\documentclass[journal=jacsat,manuscript=article]{achemso}

\usepackage[version=3]{mhchem} % Formula subscripts using \ce{}
\usepackage{comment}
\usepackage[hidelinks]{hyperref}
\usepackage{xcolor}
\usepackage{siunitx}
\usepackage{bm}
\usepackage{amsmath,amssymb}
\usepackage{xr}
\mciteErrorOnUnknownfalse

\DeclareUnicodeCharacter{03C0}{\ensuremath{\pi}}
\newcommand{\didv}{{d}\textit{I}/{d}\textit{V} }

\makeatletter

\author{Elia Turco}
\affiliation[Empa]{nanotech@surfaces Laboratory, Empa---Swiss Federal Laboratories for Materials Science and Technology, 8600 D\"{u}bendorf, Switzerland}
\email{elia.turco@empa.ch}
\author{Fupeng Wu}
\affiliation[Max]{Max Planck Institute of Microstructure Physics Weinberg 2, 06120 Halle, Germany}
\alsoaffiliation[CFAED]{Center for Advancing Electronics Dresden (cfaed) \& Faculty of Chemistry and Food Chemistry, Technische Universität Dresden, Mommsenstrasse 4, 01062 Dresden, Germany}
\author{Gon\c{c}alo Catarina} 
\affiliation[Empa]{nanotech@surfaces Laboratory, Empa---Swiss Federal Laboratories for Materials Science and Technology, 8600 D\"{u}bendorf, Switzerland}
\author{Nils Krane}
\affiliation[Empa]{nanotech@surfaces Laboratory, Empa---Swiss Federal Laboratories for Materials Science and Technology, 8600 D\"{u}bendorf, Switzerland}
\author{Ji Ma}
\affiliation[Max]{Max Planck Institute of Microstructure Physics Weinberg 2, 06120 Halle, Germany}
\alsoaffiliation[CFAED]{Center for Advancing Electronics Dresden (cfaed) \& Faculty of Chemistry and Food Chemistry, Technische Universität Dresden, Mommsenstrasse 4, 01062 Dresden, Germany}
\author{Roman Fasel}
\affiliation[Empa]{nanotech@surfaces Laboratory, Empa---Swiss Federal Laboratories for Materials Science and Technology, 8600 D\"{u}bendorf, Switzerland}
\alsoaffiliation[University of Bern]{Department of Chemistry, Biochemistry and Pharmaceutical Sciences, University of Bern, 3012 Bern, Switzerland}
\author{Xinliang Feng}
\affiliation[Max]{Max Planck Institute of Microstructure Physics Weinberg 2, 06120 Halle, Germany}
\alsoaffiliation[CFAED]{Center for Advancing Electronics Dresden (cfaed) \& Faculty of Chemistry and Food Chemistry, Technische Universität Dresden, Mommsenstrasse 4, 01062 Dresden, Germany}
\email{xinliang.feng@tu-dresden.de}
\author{Pascal Ruffieux}
\affiliation[Empa]{nanotech@surfaces Laboratory, Empa---Swiss Federal Laboratories for Materials Science and Technology, 8600 D\"{u}bendorf, Switzerland}
\email{pascal.ruffieux@empa.ch}

\title{ Magnetic Excitations in Ferromagnetically Coupled Spin-1 Nanographenes}

\begin{document}
%%%%%%%		Abstract			%%%%%%%    
\begin{abstract}
In the quest for high-spin building blocks to form covalently bonded 1D or 2D materials with controlled magnetic interactions, $\pi$-electron magnetism provides an ideal framework to engineer large ferromagnetic interactions between nanographenes. As a first step in this direction, we investigate the spin properties of ferromagnetically coupled triangulenes, triangular nanographenes with spin $S = 1$. Combining in-solution synthesis of rationally designed molecular precursors and on-surface synthesis, we achieve covalently bonded $S = 2$ triangulene dimers and $S = 3$ trimers on Au(111). Starting from the triangulene dimer, we thoroughly characterize its low-energy magnetic excitations using inelastic electron tunneling spectroscopy (IETS). IETS reveals conductance steps identified as a quintet to triplet excitation, and a zero-bias peak stemming from higher-order spin-spin scattering of the 5-fold degenerate ferromagnetic ground state. The Heisenberg picture captures the relevant parameters of inter-triangulene ferromagnetic exchange, and its successful extension to the larger $S = 3$ system confirms the model's accuracy. We expect that the addition of ferromagnetically coupled building blocks to the toolbox of magnetic nanographenes opens new opportunities to design carbon materials with complex magnetic ground states.
\end{abstract}

%%%%%%%%%%%%%%%%%%%%%%%%%%%%%%%%%%%%%%%%%%%%%%%%%%%%%%%%%%
%%%%%%%%%%%%%%%%%%%%%%%%%%%%%%%%%%%%%%%%%%%%%%%%%%%%%%%%%%
%%%%%%%%%%%%%%%%%%%%%%%%%%%%%%%%%%%%%%%%%%%%%%%%%%%%%%%%%%

%%%%%%%%%%%%%%%%%%%%%%%%%%%%%%%%%%%%%%%%%%%%%%%%%%%%%%%%%%
%%%%%%%%%%%%%%%%%%%%%%%%%%%%%%%%%%%%%%%%%%%%%%%%%%%%%%%%%%
%%%%%%%%%%%%%%%%%%%%%%%%%%%%%%%%%%%%%%%%%%%%%%%%%%%%%%%%%%

%%%%%%%		 Main Text			%%%%%%% 

%	For Communications for Angewandte Chemie, please remove headlines for Introduction, Results and Discussion and Conclusion

\section{Introduction}
\label{introduction}
Ferromagnetic exchange is the dominant form of interaction in atomic magnets, where electrons within a given shell tend to maximize the total spin quantum number $S$\cite{hund_zur_1925}. It can be regarded as the antithesis of bonding, hence rare to find in organic molecules \cite{rajca_organic_1994}. The challenge in synthesizing high-spin organic molecules is to tune the symmetry properties of the unpaired electrons' wavefunction, making the parallel spin alignment energetically favorable. The fundamental interest in high-spin organic molecules and their wide range of prospective applications (spin qubits, optoelectronics and spintronics) \cite{tan_electronic_2022, coronado_molecular_2020, chen_quantum_2024} boosted tremendous progress in this field, witnessed by the synthesis of stable high-spin organic molecules, clusters, and polymers\cite{rajca_physical_2005, shu_stable_2023, gallagher_high-spin_2015,zhang_high-spin_2022}. The main bottleneck in the synthesis of these materials is the weak magnetic exchange coupling between the high-spin units, limiting their application to very low temperatures\cite{ratera_playing_2012}.

\begin{figure}[h]
\begin{center}
\includegraphics[width=8.5cm]{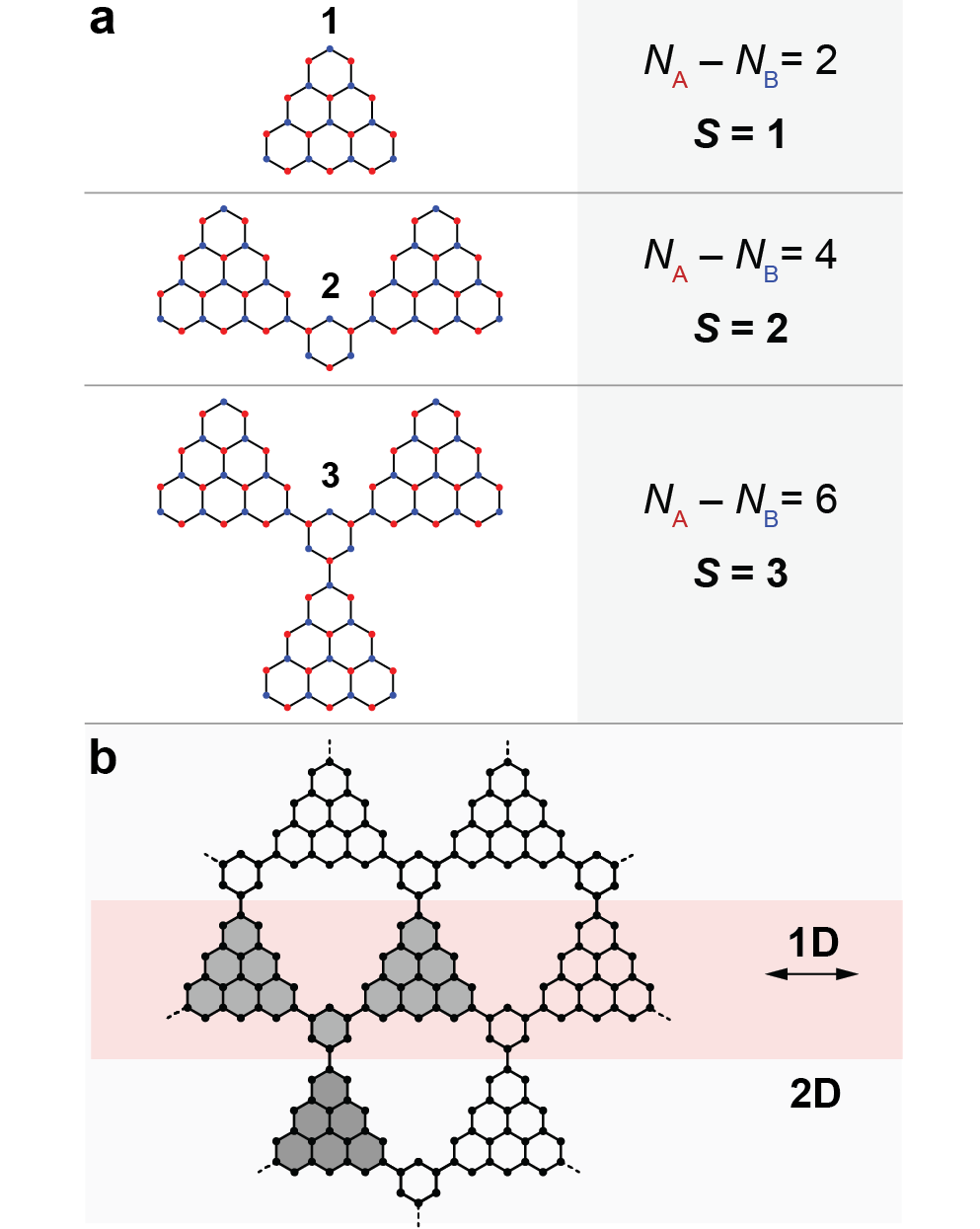}
\caption{Ferromagnetic coupling of triangulene building blocks. 
a) Coupling two and three triangulene units via 1,3- and 1,3,5-phenylene spacers leads to structures \textbf{2} and \textbf{3}, respectively. 
Ovchinnikov's counting rule predicts a total spin quantum number $S$ according to the sublattice imbalance (red and blue filled circles denote the two sublattices). b) Structures \textbf{2} and \textbf{3} can be regarded as prototypical high-spin building blocks towards the fabrication of 1D and 2D ferromagnetic materials.}
\label{structures}
\end{center}
\end{figure}
 
In this regard, open-shell nanographenes offer the possibility of achieving strong intermolecular exchange correlations through covalent coupling of the spin units\cite{turco_-surface_2021, mishra_large_2021, krane_exchange_2023}. 
Taking advantage of the bipartite honeycomb lattice of graphene, it is possible to design nanographenes (NGs) hosting unpaired $\pi$-electrons\cite{oteyza_carbon-based_2022}, with a total spin $S$ given by Ovchinnikov's counting rule $S = |N_A -N_B|/2$, where $N_A$ and $N_B$ are the number of carbon atoms in the two sublattices \cite{Lieb1989,ovchinnikov_multiplicity_1978}.
The family of zigzag-edged triangular NGs, commonly denoted as [n]triangulenes, carry a total spin $S$ that scales with the triangulene size as $S=(n-1)/2$, which makes them well suited as magnetic building blocks\cite{su_triangulenes_2020}.
\begin{scheme*}[h]
\begin{center}
\includegraphics[width=15.0cm]{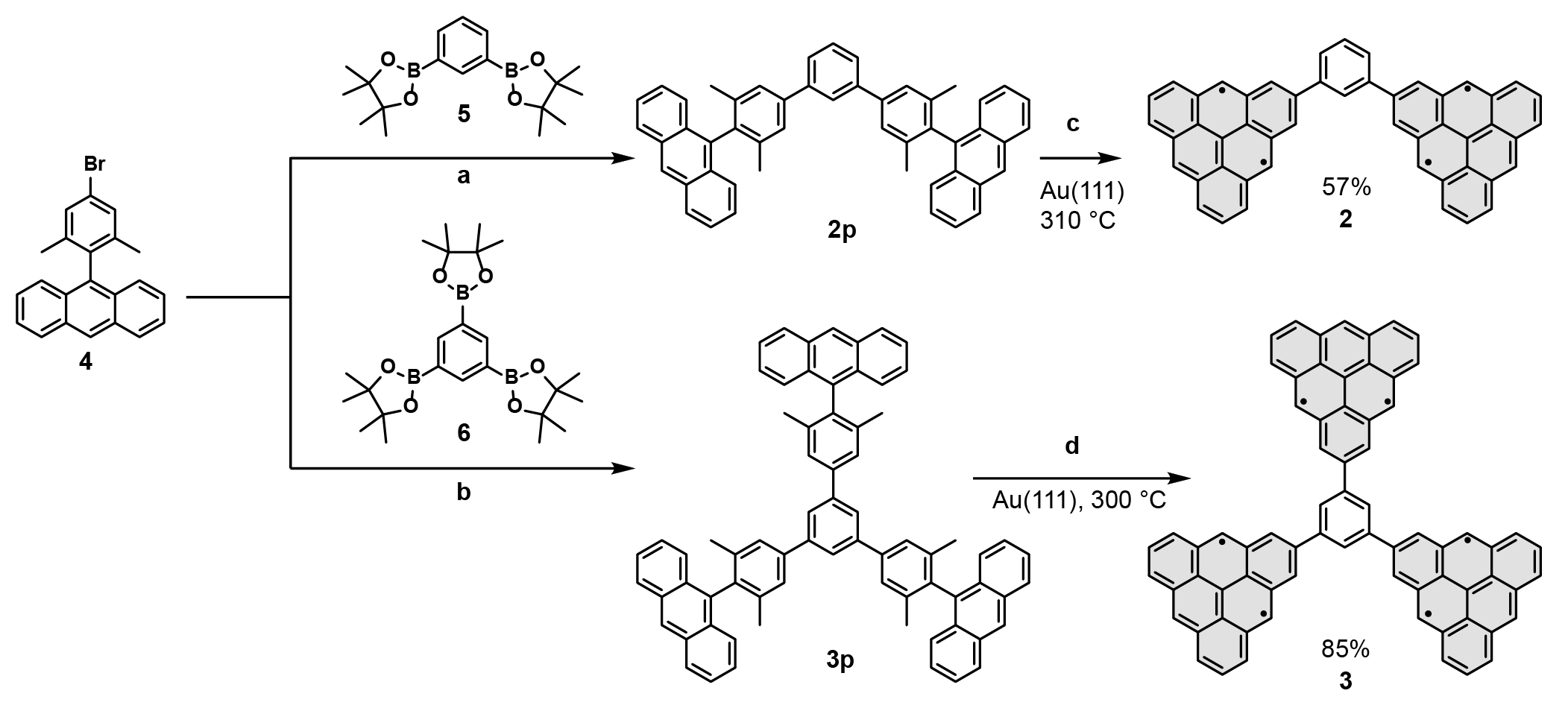}
\caption{Synthetic route to ferromagnetic triangulene dimer \textbf{2} and trimer \textbf{3}. Reagents and conditions: (a) PdCl$_{2}$(dppf)CH$_{2}$Cl$_{2}$, K$_{3}$PO$_{4}$, dioxane, 100 °C, 16 h, 34\%. (b) PdCl$_{2}$(dppf)CH$_{2}$Cl$_{2}$, K$_{3}$PO$_{4}$, dioxane, 85 °C, 16 h, 65\%. (c,d) respectively  Au (111) held at 310 ° C and 300 ° C. }
\label{solution synthesis}
\end{center}
\end{scheme*}

\begin{figure*}[h]
\begin{center}
\includegraphics[width=17.3cm]{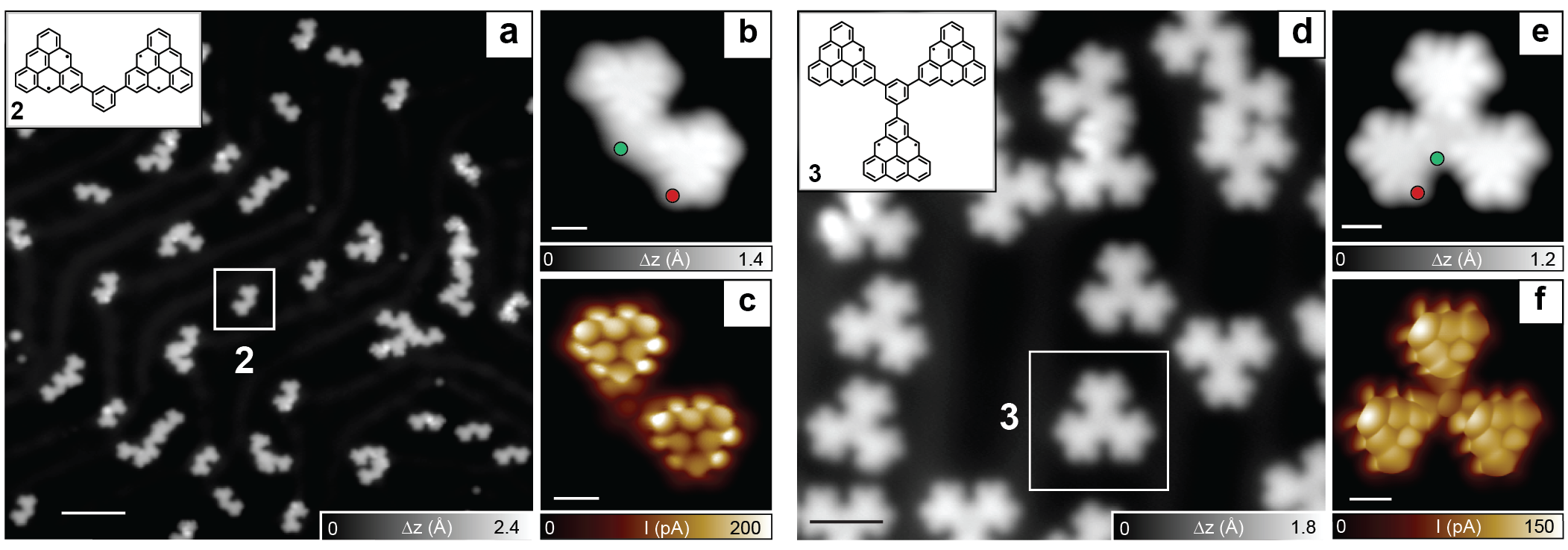}
\caption{On-surface synthesis and structural characterization of structures \textbf{2} and \textbf{3}. (a,d) Overview STM images after the deposition and annealing step of \textbf{2p} and \textbf{3p} on Au(111) ($V = -0.1$ V, $I = 100 $pA). Target structures \textbf{2} and \textbf{3} are highlighted with squares. The corresponding STM images (b,e) ($V = -0.6$ V, $I = 100 $pA) and bond-resolved STM images (c,f) (acquired with a carbon monoxide functionalized tip and tunneling parameters: $V = -5 $ mV, $I = 50$ pA, $\Delta h=-0.6 $\AA  ) demonstrate the successful synthesis of \textbf{2} and \textbf{3}.}
\label{on-surface synthesis}
\end{center}
\end{figure*}
\noindent$[n]$triangulenes contain $[n-1]$ zero energy modes hosted on the majority sublattice, subject to large Coulomb repulsion and therefore strong intra-triangulene ferromagnetic exchange ($J_{FM} >$ 100 meV)\cite{ortiz_exchange_2019}.

The $S = 1$ triangulene (\textbf{1} in Figure \ref{structures}a) has been the subject of several experimental studies, focusing mainly on its fundamental magnetic properties\cite{turco_observation_2023,pavlicek_synthesis_2017} and the correlated magnetic ground states emerging from antiferromagnetic coupling of the $S = 1$ building blocks\cite{mishra_collective_2020,mishra_observation_2021,hieulle_-surface_2021,calvo-fernandez_multi-orbital_2023,su_-surface_2021}.

\noindent In contrast to the antiferromagnetic case, only a few examples of ferromagnetically coupled NGs have been reported so far\cite{li_uncovering_2020, su_atomically_2020, cheng_-surface_2022, zheng_designer_2020, du_orbital-symmetry_2023}. Furthermore, an in-depth understanding of the magnetic excitations of degenerate ground states is still lacking.
With this aim, we examine the ferromagnetic coupling of two and three $S = 1$ triangulenes, here realized with the corresponding NG structures \textbf{2} and \textbf{3} (Figure \ref{structures}a). They can be regarded as prototypical spin clusters, conceived to characterize the relevant spin Hamiltonian parameters and eventually employ \textbf{2} and \textbf{3} as building blocks for ferromagnetic 1D chains and 2D networks (Figure \ref{structures}b).

\begin{figure*}[h]
\begin{center}
\includegraphics[width=17.4cm]{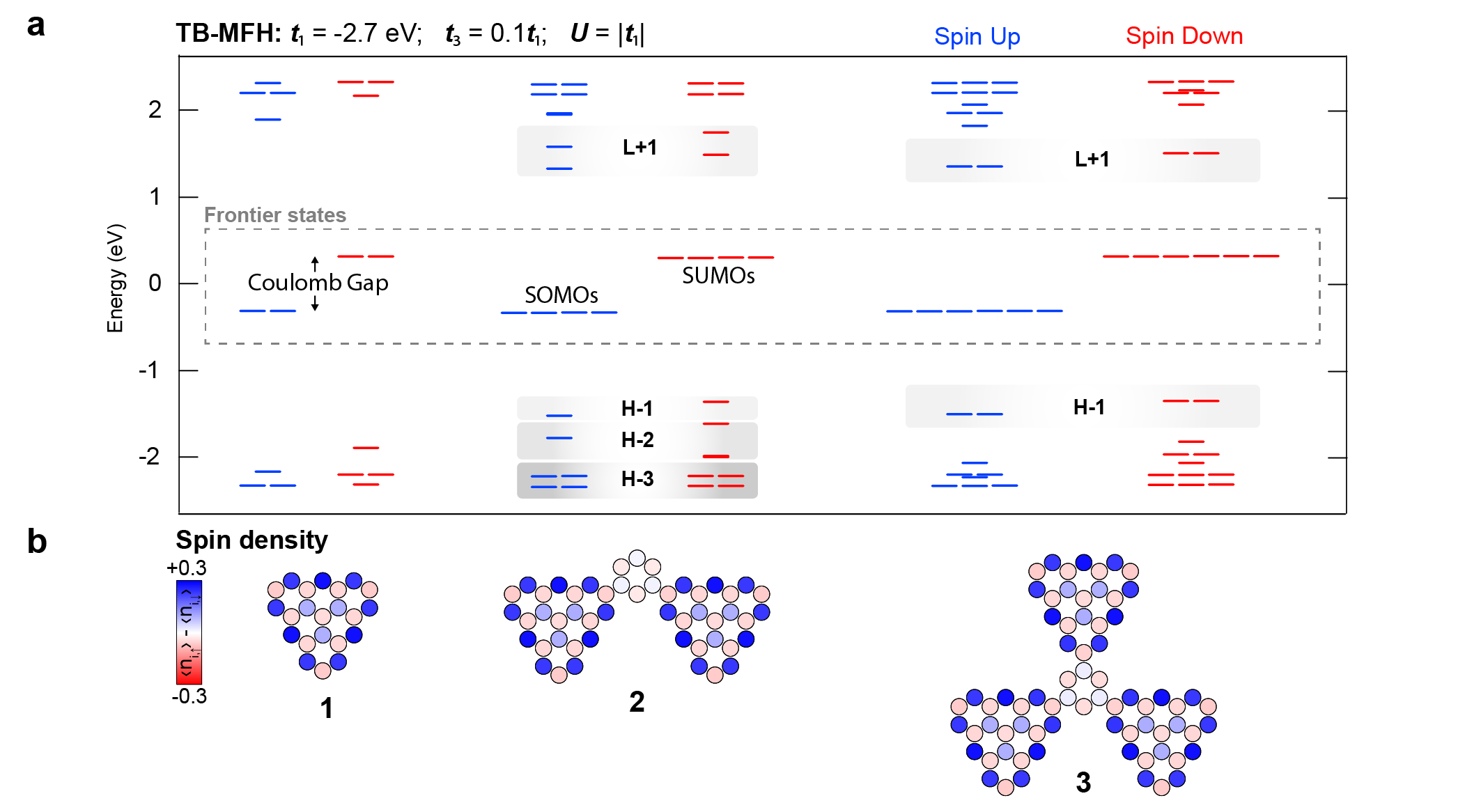}
\caption {TB-MFH-simulated electronic and magnetic properties of \textbf{1}, \textbf{2} and \textbf{3}. (a) Energy spectra calculated using third-neighbor hopping $t_3 = 0.1 t_1$ and on-site Coulomb repulsion $U = |t_1|$. (b) Corresponding spin density plots, where blue and red filled circles denote the populations of spin-up and spin-down electrons.}
\label{MFH}
\end{center}
\end{figure*}
The synthetic approach used here consists of an in-solution synthesis of the precursor molecules \textbf{2p} and \textbf{3p} and on-surface synthesis of \textbf{2} and \textbf{3} upon thermal annealing on a Au(111) surface. By Combining STS, STM and theoretical calculations, we then elucidated the spectroscopic features of \textbf{2}'s and \textbf{3}'s high-spin ground states.

\section*{Results And Discussion}

Triangulene dimer \textbf{2} and triangulene trimer \textbf{3} were synthesized via a combined in-solution and on-surface synthesis approach (see Scheme \ref{solution synthesis}). First, the key building block 9-(4-bromo-2,6-dimethylphenyl)anthracene \textbf{4} was synthesized through a five-step procedure according to our previous work \cite{mishra_collective_2020}. Then, Suzuki-coupling of \textbf{4} with commercially available 1,3-bis(4,4,5,5-tetramethyl-1,3,2-dioxaborolan-2-yl)-benzene \textbf{5} or 1,3,5-tris(4,4,5,5-tetramethyl-1,3,2-dioxaborolan-2-yl)benzene \textbf{6} gave precursor \textbf{2p} and \textbf{3p} in 34\% yield and 65\% yield, respectively.

The target structures \textbf{2} and \textbf{3} in Scheme \ref{structures} were achieved by on-surface synthesis starting from molecular precursors \textbf{2p} and \textbf{3p}, respectively.
For the synthesis of \textbf{2}, a submonolayer of \textbf{2p} was deposited onto an atomically clean Au(111) surface at room temperature (see SI Figure S4a) and subsequently annealed at 310 °C to promote oxidative ring closure of the methyl groups. STM imaging in Figure \ref{on-surface synthesis}a reveals the presence of a majority of isolated dimers along with covalently bonded molecular clusters. By an accurate inspection of various STM images, we found that 57\% of the isolated dimers show a uniform and symmetric topography consistent with \textbf{2} (Figure \ref{on-surface synthesis}b). Bond-resolved (BR) STM imaging (Figure \ref{on-surface synthesis}c) further corroborates this assignment. 
Similarly, \textbf{3} is obtained from the precursor \textbf{3p} deposited onto a clean Au(111) surface held at room temperature (Figure S4b) or at 300 °C (Figure \ref{on-surface synthesis}d).  
The latter preparation yielded mainly covalently bonded molecular clusters, but 85\% of the isolated molecules feature a uniform clover-like topography shown in Figure \ref{on-surface synthesis}e characteristic of structure \textbf{3} and demonstrated by BR-STM imaging in Figure \ref{on-surface synthesis}f.

\vspace{10pt}
\noindent
\textbf{Electronic Structure}

To establish the basic theoretical concepts relevant to the electronic and magnetic properties of the triangulene-based nanographenes, we first resort to Tight-Binding (TB) level of theory, where electron correlations are considered through the Mean Field Hubbard (MFH-TB) approximation. Figure \ref{MFH}a shows the energy diagram calculated with first and third neighbor hopping TB-MFH model for the [3]triangulene monomer (1), the ferro dimer (2), and ferro trimer (3). As follows from the sublattice imbalance, each triangulene unit hosts two zero energy states (ZES), mainly residing at the zigzag edge and on the same sublattice of the NG \cite{sutherland_localization_1986, fernandez-rossier_magnetism_2007, yazyev_emergence_2010}, giving rise to strong intra-triangulene Hund exchange. Once the on-site Coulomb repulsion is considered, these states split into singly occupied (unoccupied) molecular orbitals SOMOs (SUMOs) separated in energy by a Coulomb gap. When connecting the triangulene units through the meta positions of a benzene ring, such as 1,3- or 1,3,5-
phenylene spacers,  the four or six ZES live on the same sublattice and if we consider a third neighbor hopping $t_3 \neq 0$, Hund exchange provides the expected inter-triangulene ferromagnetic coupling\cite{jacob_theory_2022}. Accordingly, \textbf{2} and \textbf{3} are expected to host, respectively, a $S = 2$ and $S = 3$ ground state with the expected spin distributions reported in Figure \ref{MFH}b. 

\begin{figure*}[h]
\begin{center}
\includegraphics[width=17.4cm]{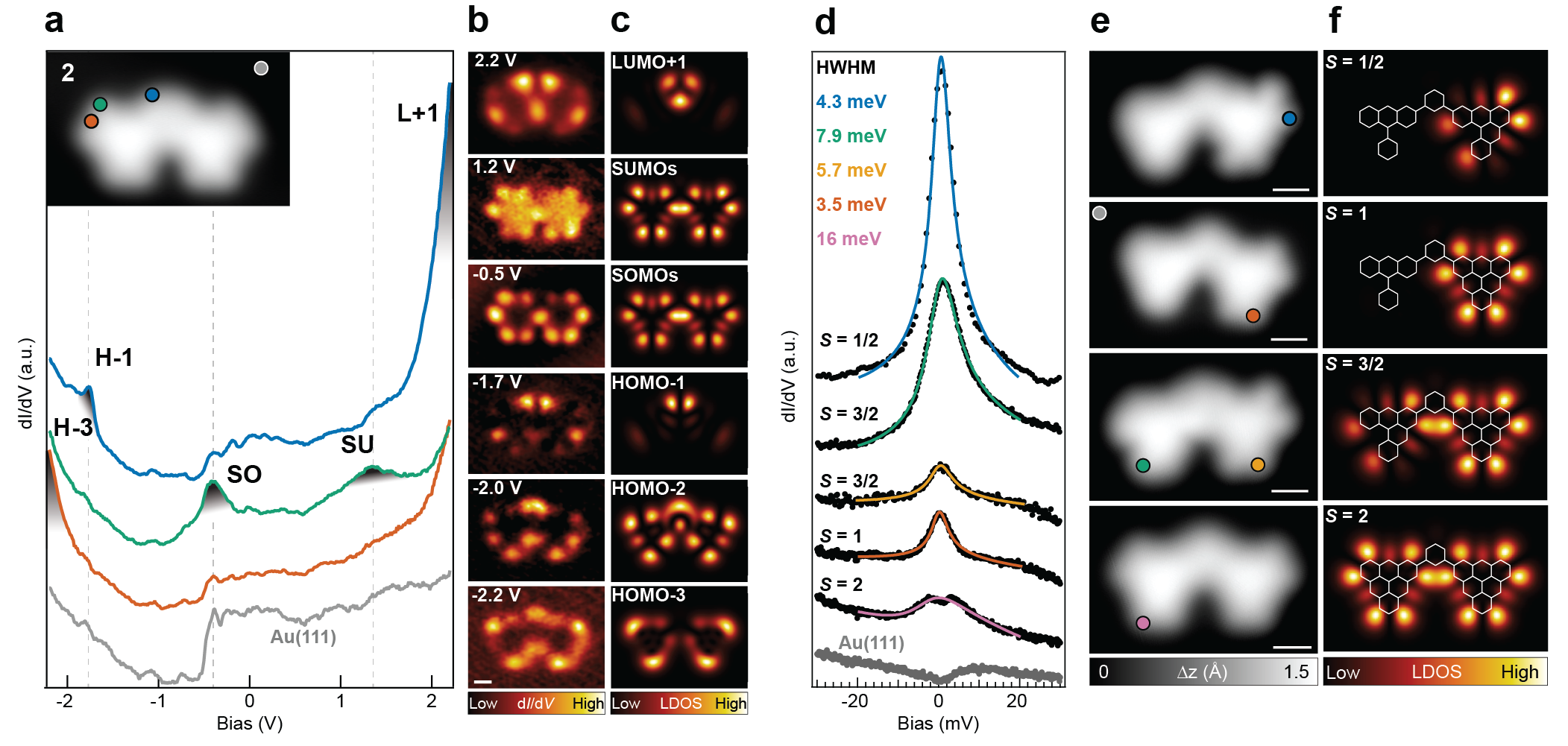}
\caption{Electronic and magnetic characterization of \textbf{2}. (a) \didv spectra acquired with a metal tip at three different locations (marked with filled circles) on \textbf{2}, revealing five distinct molecular orbital resonances (MO) (open feedback parameters: $V=-2.2$ V, $I=350$ pA; $V_{rms}=14 $mV). (b) Constant-current \didv spatial maps of the detected MO resonances ( $I=350$ pA; $V_{rms}=14 $  mV), along with the corresponding MFH-LDOS maps (c). (d) Low-bias \didv spectra of \textbf{2} and its dihydro-intermediates ($V_{rms}\leq 1.4 $mV), achieved through a tip-induced step-by-step dehydrogenation of the very same molecule. The HWHM of each zero-bias resonance was extracted by Frota fitting\cite{frota_shape_1992} and the values are reported in the inset. At each manipulation step, the corresponding low-bias ( $V=-0.1$ V, $I=50-100$ pA) STM image (e) is compared to the MFH-LDOS map (f) of the spin-carrying orbitals of the intermediate structures.}
\label{electronic properties}
\end{center}
\end{figure*} 
We now focus on the hitherto unexplored electronic properties of structure \textbf{2} on Au(111), while the electronic characterization of \textbf{3} is reported in the Supplemental Material \cite{SI}. 
In Figure \ref{electronic properties}a the STS measurements of an isolated \textbf{2} molecule on Au (111) are shown. The
differential conductance d\textit{I}/d\textit{V} spectra reveal the presence of five distinct resonances located at different positions on the molecule. To investigate how these resonances of the local density of states (LDOS) extend over the molecule, we carried out constant-current d\textit{I}/d\textit{V} maps at each resonant energy, shown in Figure \ref{electronic properties}b. A direct comparison with the calculated TB-MFH LDOS maps (shown in Figure \ref{electronic properties}c) allowed proper labeling of the experimentally observed resonances. As expected, the frontier orbitals are singly occupied (unoccupied) electronic states sharing the very same LDOS, separated in energy by a Coulomb gap of 1.7 eV. The latter argument, bolstered by the perfect match of the calculated LDOS with the experimental resonances, validates our initial assumption on the $S = 2$ ground state of \textbf{2}.

\vspace{4pt}

\noindent
\textbf{Magnetic Characterization}

We now shift our focus to the low-energy manifestations of \textbf{2}'s magnetic ground state. 
%Up to now most of the studies on coupled magnetic NGs were conducted on non degenerate $S = 0$ systems, while here 
The coupling design via  \textit{meta}-junctions is predicted to lead to weak ferromagnetic interactions, based on a combination of ferromagnetic Hund exchange (via $t_3$) and Coulomb-driven superexchange mechanisms\cite{jacob_theory_2022}.
It is also well known that electrons of a nearby metal surface, here Au(111), have the tendency to screen unpaired electrons\cite{kondo_resistance_1964} leading to a zero-bias resonance in the conductance spectrum\cite{turco_observation_2023}, and if any low-energy spin excitations are present, replicas of the zero-energy peak are expected at the excitation thresholds \cite{ternes_spin_2015}. 
Contrary to antiferromagnetically coupled systems, the $S = 2$ ground state of \textbf{2} should feature both spin excitation steps and a Kondo peak \cite{kondo_resistance_1964} due to higher-order spin-spin scattering of the degenerate ground state. To better resolve these low-energy properties, we map the low-bias d\textit{I}/d\textit{V} spectrum as a function of the total spin \textit{S} using step-by-step tip-induced dehydrogenation of a dimer molecule where the four spins were initially absent due to hydrogen passivation (more details in SI). Figures \ref{electronic properties}d,e show respectively the d\textit{I}/d\textit{V} spectrum and the STM image after each manipulation step. The correct assignment of the ground state $S$ to the respective chemical structure (Figure \ref{electronic properties}f)  is corroborated by the calculated MFH-LDOS maps of the frontier MOs shown in panel (f), where the presence of a dihydro group is taken into account in the model by removing the corresponding carbon atom. Both $S = \frac{1}{2}$ and $S = 1$ intermediates feature a Kondo peak, albeit with distinct relative amplitudes. This observation is consistent with previous findings\cite{turco_observation_2023, nevidomskyy_kondo_2009, jacob_renormalization_2021}. The asymmetric spin system spin$\frac{1}{2}$-spin1 ($S=\frac{3}{2}$) shows a significantly wider resonance compared to cases $S=\frac{1}{2}$ and $S = 1$. This observation may indicate the coexistence of low-energy magnetic excitations and a zero-energy peak; however, these energy scales are challenging to resolve at the experimental temperature of 4.5$~$K and therefore subject to future analysis. To quantitatively evaluate these differences, the half width at half maximum (HWHM) of each zero bias resonance was extracted by fitting the measured lineshape with a Frota function\cite{frota_shape_1992} and the corresponding values are reported in the inset of Figure \ref{electronic properties}d. Unlike the dimer intermediates, it is evident that the $S = 2$ d\textit{I}/d\textit{V} spectrum cannot be fitted with a single Frota, therefore requiring a more accurate analysis of its magnetic properties, as described below.

\begin{figure*}[h]
\begin{center}
\includegraphics[width=17.4cm]{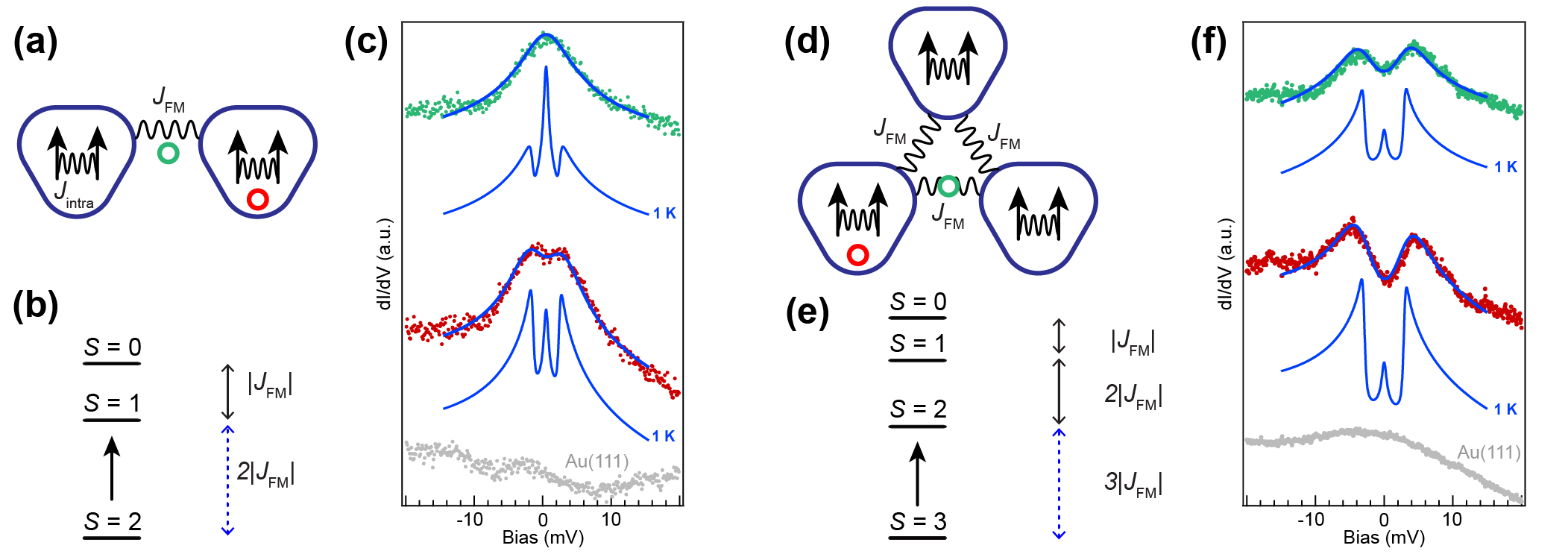}
\caption{Comparing the magnetic properties of \textbf{2} and \textbf{3}. (a,d) Heisenberg models of \textbf{2} and \textbf{3} are depicted, where each triangulene unit is considered as $S = 1$ unit. (b,e) Corresponding energy level scheme obtained by analytical solution of the Heisenberg model and confirmed by CAS calculations. (c,f) High-resolution \didv spectroscopy acquired respectively on \textbf{2} and \textbf{3} on two distinct locations for each molecule, depicted as coloured filled circles in Figure \ref{on-surface synthesis} (b,e). By including the spin-spin scattering processes (perturbative part) into the Heisenberg Hamiltonian and solving to the third order, we could nicely fit the measured \didv spectra and extract the effective ferromagnetic exchange, which was found to be approximately 1 meV for both \textbf{2} and \textbf{3}. Open feedback parameters \didv spectra: (c) $V=-30$ mV, $I=600$ pA;  (f) $V=-20$ mV, $I=500$ pA. Lock-in modulation (c) $V_{rms}=350 $ $\mu$V, (f) $V_{rms}=700 $$\mu$V.  }

\label{Low-bias spectroscopy}
\end{center}
\end{figure*} 

Figure \ref{Low-bias spectroscopy}c shows high-resolution d\textit{I}/d\textit{V} spectra highlighting the low-energy magnetic excitations measured at two distinct locations on \textbf{2} with a metal tip. It is apparent the qualitative difference of the spectrum acquired on the two locations on the molecule. Although the spectrum acquired on the central part of the molecule (green spectrum) appears to be a broad zero-bias peak, the one taken on the outer side of the triangulene unit (red spectrum) clearly reveals the presence of two peaks at about $\pm2$ meV. As shown in Figure S5 (SI), very similar low-energy features are inherently present in the STS spectrum if CO-functionalized are used\cite{vitali_surveying_2010} (because of inelastic vibrational excitations); therefore, to avoid erroneous interpretations, in the following analysis only metal tips will be considered.

To shed light into these elusive magnetic features and deepen our understanding of \textbf{2}, we first resort to theory by considering respectively six and eight of the TB single particle states closer to Fermi and solving the many-body Hubbard Hamiltonian exactly in this restricted basis, commonly known as CAS(6,6) and CAS(8,8) calculations. We find both calculations in line with the expected magnetic spectrum, with the triplet and singlet excited states, respectively, $2|J_{FM}|$ and $3|J_{FM}|$ away from the quintet ground state (excitation spectrum in Figure \ref{Low-bias spectroscopy}b). %Although the calculated $|J_{FM}|$ value strongly depends on the chosen Coulomb repulsion term, our calculations show a significantly reduced $|J_{FM}|$ when considering HOMO-2 and LUMO+2 in the CAS(8,8) calculations, as recently shown in Ref. \cite{jacob_theory_2022}. 
On this basis, we can now represent \textbf{2} with a simple Heisenberg Hamiltonian (Figure \ref{Low-bias spectroscopy}a) were two spin-1 units $\Vec{S_1}$ and $\Vec{S_2}$ are coupled by an effective ferromagnetic interaction $J_{FM}\Vec{S_1}\cdot \Vec{S_2}$. By including a tunnel junction and the corresponding spin-flip scattering processes up to the third order, we obtain the calculated d\textit{I}/d\textit{V} intensity\cite{ternes_probing_2017} (here depicted by the blue spectra). By fitting the d\textit{I}/d\textit{V} spectrum in red (Figure \ref{Low-bias spectroscopy}c), a $|J_{FM}|$ value of 0.98 meV is obtained (more details on the fitting procedure can be found in the Supplementary file). To better visualize the inherent magnetic excitations, we recalculated the resulting fit at an effective temperature of 1K and reported the results below. The latter reveals two distinct features: a sharp zero energy resonance, the result of third-order spin scattering processes of the degenerate ground state, and two symmetric peaks at about $\pm 2 $ meV, due to the spin-flip transition between the $S = 2$ ground state and the $S = 1$ excited state (allowed by spin selection rules). If we now fix $|J_{FM}|$ and try to fit the green spectrum acquired in the center of \textbf{2}, the fit does not converge. Indeed, when the tip is placed in between the two spin units, third-order spin scattering processes can occur simultaneously on both units, as denoted by the increased zero-energy peak with respect to the spin excitation peaks, as shown in the calculated 1 K spectrum. By including this effect, we obtain a perfect match with the experimental data, while keeping the very same Kondo scattering term $J_\rho$ and effective temperature $T_{eff}$.

With the elucidation of the mechanisms underlying \textbf{2}'s low-energy magnetic excitations, we are now able to extend the analysis to \textbf{3} and eventually compare the two systems.
The computationally challenging CAS(10,10) corroborates the ferromagnetic solution for \textbf{3}, with three excited states defined by spin quantum numbers $S = 2$, $S = 1$ and $S = 0$, separated in energy from the septet ($S = 3$) ground state, respectively $3|J_{FM}|$, $5|J_{FM}|$ and $6|J_{FM}|$, as depicted in the energy scheme in Figure \ref{Low-bias spectroscopy}e. The d\textit{I}/d\textit{V} spectra measured on \textbf{3} and reported in  Figure \ref{Low-bias spectroscopy}f unveil the presence of low-energy magnetic excitations ascribed to the septet-quintet transition, while excitations from the ground state to higher energy states are forbidden by spin selection rules. To fit the experimental data and extract $|J_{FM}|$, we resort to the previously adopted Heisenberg model (Figure \ref{Low-bias spectroscopy}d). Using the very same scattering parameters and $|J_{FM}|$ adopted in \textbf{2} an excellent fit of the red spectrum in Figure \ref{Low-bias spectroscopy}f was obtained. By placing the tip in between two spin-1 units, an increased intensity of the zero-energy peak is recorded, due to the simultaneous scattering of the two distinct spin units.
The in-depth analysis of \textbf{2} and \textbf{3} considered in this work highlights the relevance of high-order spin scattering effects in the low-energy magnetic behavior of systems with degenerate ground states. We also demonstrated the accuracy of our approach by successfully extending \textbf{2}'s Hamiltonian parameter space to the more complicated $S = 3$ system (labelled as \textbf{3}), thus widening the future implications of our work to extended ferromagnetically coupled spin systems. 

\section*{Conclusion}
\label{conclusion}
	
This study shows the successful on-surface synthesis of high-spin NGs where triangulene spin-1 units are ferromagnetically coupled via 1,3- and 1,3,5-phenylene spacers into dimers and trimers. The considered bonding scheme is predicted to create magnetic NGs with a total spin quantum number $S$ equal to the number of connected triangulene units. By high-resolution STS we carried out a comprehensive electronic characterization of both systems, revealing the presence of low-energy magnetic excitations. By comparing our results with different levels of theory, we proved the open-shell $S=2 $ and $S =3$ ground state of respectively the dimer and trimer, ascribing the observed low-energy features to quintet-triplet and septet-quintet spin excitations. Despite the low ferromagnetic exchange coupling intrinsic to the employed bonding scheme, our analysis showed the important role of quantum correlations embodied by the observed zero-energy peaks.     

\section*{Acknowledgements}
This research was financially supported by the EU Graphene Flagship (Graphene Core 3, 881 603), ERC Consolidator Grant (T2DCP, 819 698), H2020-MSCA-ITN (ULTIMATE, No. 813 036), Swiss National Science Foundation (SNF-PiMag, No. CRSII5\_205987) the Center for Advancing Electronics Dresden (cfaed), H2020-EU.1.2.2.-FET Proactive Grant (LIGHT‑CAP, 101 017 821), and the DFG-SNSF Joint Switzerland-German Research Project (EnhanTopo, No. 429 265 950).

\noindent Skillful technical assistance by Lukas Rotach is gratefully acknowledged.

\noindent E.T. and F.W. contributed equally to this work.

\section*{Conflict of Interest}

The authors declare no conflict of interest.

%%%%%%%%%%%%%%%%%%%%%%%%%%%%%%%%%%%%%%%%%%%%%%%%%%%%%%%%%%
%%%%%%%%%%%%%%%%%%%%%%%%%%%%%%%%%%%%%%%%%%%%%%%%%%%%%%%%%%
%%%%%%%%%%%%%%%%%%%%%%%%%%%%%%%%%%%%%%%%%%%%%%%%%%%%%%%%%%

%%%%%%%		References			%%%%%%% 

\setlength{\bibsep}{0.0cm}
\bibliography{references}

%%%%%%%%%%%%%%%%%%%%%%%%%%%%%%%%%%%%%%%%%%%%%%%%%%%%%%%%%%
%%%%%%%%%%%%%%%%%%%%%%%%%% Supplemental material%%%%%%%%%%
\setcounter{figure}{0}    
\renewcommand{\thefigure}{S\arabic{figure}}
\section{Experimental Methods}

\subsection{Sample Preparation and Scanning Probe Measurements}
\vspace{8pt}

STM measurements were performed with a commercial low-temperature STM/AFM from Scienta Omicron operated at a temperature of $4.5$ K and a base pressure below $5 \cdot 10^{-11}$ mbar. The Au(111) single crystal surfaces were prepared by iterative Ar$^+$ sputtering and annealing cycles. Before sublimation of molecules, the surface quality was verified through STM imaging. The powders of \textbf{2p} and \textbf{3p} precursors were filled into quartz crucibles of a home-built evaporator and sublimed at 280 ° C and 290 ° C, respectively, on the surfaces of the single crystal. STM images were acquired in both constant-current (overview and high-resolution imaging) and constant-height (bond-resolved imaging) modes, \didv spectra were acquired in constant-height mode, and \didv maps were acquired in constant-current mode. Indicated bias voltages are given with respect to the sample. Unless otherwise noted, all measurements were performed with metallic tips. Differential conductance \didv spectra and maps were obtained with a lock-in amplifier. Modulation voltages (root mean square amplitude $V_{rms}$) for each measurement are provided in the respective figure caption. Bond-resolved STM images were acquired in constant-height mode with CO-functionalized tips at low bias voltages while recording the current signal. Open feedback parameters on the molecular species and the subsequent lowering of the tip height ($\Delta z$) for each image are provided in the respective figure captions. The data were processed with Wavemetrics Igor Pro software.
\subsection{Hydrogen passivation }
In figure 4 (d,e) of the main text, dihydro intermediates of \textbf{2} are reported. These structures are formed by passivation of \textbf{2}'s \textit{active } spin sites by hydrogen diffusion on the metal surface, subsequently to the cyclodehydrogenation reaction step. Therefore, it naturally occurs to find structures where the unpaired electrons are partially or fully quenched by dihydro groups, thus allowing the detection and sequential manipulation of intermediates with various spin ground states \textit{S}. More details on the tip-based manipulation method can be found in Ref. \cite{turco_observation_2023}.

\subsection{Determining the yield of the target compounds} 
The reported \textbf{2's} and \textbf{3's} yields were evaluated by assessing the percentage of target molecules (with flat and uniform topography) compared to the total number of single molecules. For each system, various overview STM images were analyzed and the statistics were computed on a sample larger than a hundred of molecules.

\newpage

\section{Computational Methods}

\subsection{Tight-binding and mean-field Hubbard calculations}

TB-MFH calculations were performed by numerically solving the mean-field Hubbard Hamiltonian with third-nearest-neighbor hopping.

The corresponding Hamiltonian reads as
\begin{equation}
\mathcal{\hat{H}}_\mathrm{MFH} = \sum_{j} \sum_{\langle \alpha,\beta \rangle j,\sigma } t_j \hat{c}^\dagger_{\alpha,\sigma} \hat{c}_{\beta,\sigma} + U \sum_{\alpha, \sigma} \langle n_{\alpha,\sigma} \rangle n_{\alpha,\Bar{\sigma}}- U \sum_{\alpha} \langle n_{\alpha,\uparrow} \rangle \langle n_{\alpha,\downarrow} \rangle  ,
\end{equation}

Here, ${c_{\alpha,\sigma}}^\dagger$ and $c{_\beta,\sigma}$  denote the spin selective ($\sigma \in {\uparrow, \downarrow} $) creation and annihilation operator at sites $\alpha$ and $\beta$, $\langle\alpha,\beta\rangle_j$ ($j={1,3}$) denotes the nearest-neighbor and third-nearest-neighbor sites for j = 1, and 3, respectively, $t_j$ denotes the corresponding hopping parameters (with $t_1$  = 2.7 eV and $t_3  = 0.1 t_1$ for nearest-neighbor and third-nearest-neighbor hopping), U denotes the on-site Coulomb repulsion, $n_{\alpha,\sigma}$ denotes the number operator, and $\langle n_{\alpha,\sigma} \rangle$ denotes the mean occupation number at site $\alpha$. Orbital electron densities, $\rho$, of the $n^{th}$-eigenstate with energy $E_n$ have been simulated from the corresponding state vector $a_{n,i,\sigma}$ by

\begin{equation}
\rho_{n,\sigma}(\Vec{r}) = \Bigg| \sum_i a_{n,i,\sigma}\phi_{2p_z}(\Vec{r}-\vec{r_i}
)\bigg|^2  ,
\end{equation}

where i denotes the atomic site index and $\phi_{2p_z}$ denotes the Slater $2p_z$ orbital for carbon.
All TB-MFH calculations presented in the manuscript were done in the third-nearest-neighbor approximation and using an on-site Coulomb term $U = \lvert t_1 \rvert$.

\newpage
\subsection{Calculation of differential conductance spectra}
First, the spin hamiltonian is constructed by considering each triangulene as a spin-1 unit and a ferromagnetic Heisenberg-like exchange $J_{inter}$ between neighboring units, while the magnetic anysotropy term and non-collinear Dzyaloshinskii–Moriya spin coupling are irrelevant for this system. Notably, if we instead consider each unit as two ferromagnetically coupled spins with $J_{intra}\gg J_{inter}$, the result is identical. The d\textit{I}/d\textit{V} spectra were simulated by introducing a perturbative term into the spin Hamiltonian, which accounts for spin-flip processes up to third order in the interaction matrix elements \cite{ternes_spin_2015}.

\begin{figure*}[h]
\begin{center}
\includegraphics[width=17.4cm]{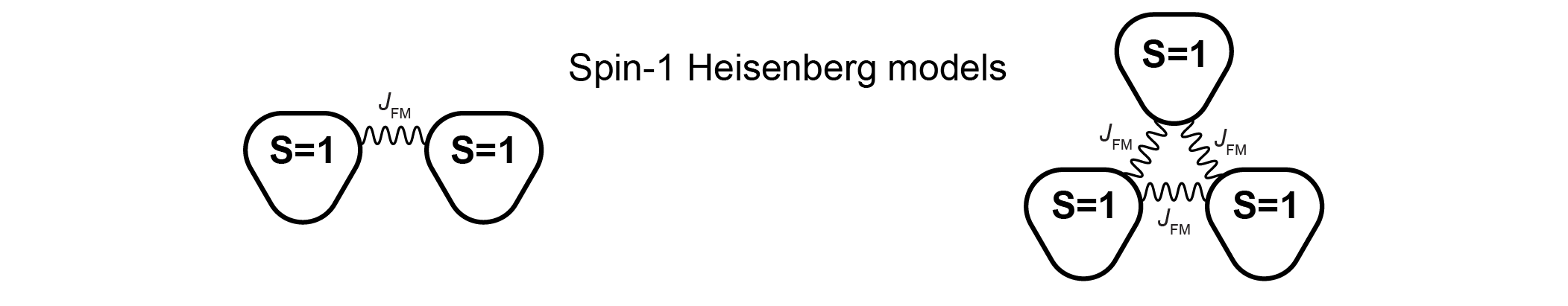}
\caption {Heisenberg model of the dimer \textbf{2} and trimer \textbf{3}.}
\label{third order calculations}
\end{center}
\end{figure*}

The data shown in Fig.4c of the manuscript have been fitted using the fitting procedures provided in ref. \cite{ternes_spin_2015}. In the following, I will briefly describe the fitting \textit{modus operandi}. \newline
\vspace{2pt}

\textbf{Frozen parameters}
\begin{itemize}
    \item {$ \omega_0 = 20$ (The bandwidth of the scattering substrate electrons in meV.}
    \item $U = 0$ (The Coulomb scattering parameter)
    \item $\eta = 0$ (The spin polarization of the tip)
    \item $V_{off} = 0$ (Voltage offset in mV)
\end{itemize}
\vspace{2pt}

\textbf{Fitting parameters}
\begin{itemize}
    \item $J_{FM} $ (Ferromagnetic Heisenberg coupling strength)
    \item {$ J \rho_s$ (The Kondo scattering parameter).}
    \item $T_0^2 $ (Tunnel barrier coupling constant, as scaling factor)
    \item $T_K$ (Effective temperature in Kelvin)
    \item $b, \sigma_0$ (respectively sloped and conductance offset) 
\end{itemize}

\textbf{Note}: First the the fitting parameters were manually set to get close to the observed spectrum, then the fitting procedure was initiated. In a second step, to accurately determine $J_{FM}$, the fitting was repeated but keeping some of the fitting parameters ($b, \sigma_0, T_0^2$) frozen.  

The spectra acquired in the external part of \textbf{2} was employed as model system to define the parameter space. To reproduce the spectrum taken in the central part of the molecule, third order scattering processes through both spin centers were considered. The fit was carried out keeping the same $J_{FM} = 0.98 $meV.

\begin{table}[h]
\centering
    \begin{tabular*}{\linewidth}{r@{\hspace{3ex}}||@{\extracolsep\fill}ccccccc}
                        & $J_{FM}$ & 
                        $J\rho_s$ &
                        $T_0^2$ & 
                        $b$ & 
                        $V_{off}$ &
                        $\sigma_0$ &
                        $T_{eff}$  \\[0.3ex]
                        \hline
        $(c) \ red$         & 0.98 & 0.19 &0.046 & -0.005 & 0.5 & 2.99 & 7.88
        \\[1ex]
        $(c) \ green$         & 0.98 & 0.25 &0.024 & -0.0002 & 0.5 & 7.03 & 7.88 
        \\[1ex]
        $(f) \ red$        & 0.98  & 0.19 &0.12 & -0.006 & 0.0 & -1.37 & 7.88 
        \\[1ex]
        $(f) \  green$       & 0.98 & 0.31 &0.055& -0.002 & 0.0 & 2.3 & 7.88      
    \end{tabular*}        
\caption{\label{tab:TK2} 
Parameters used to fit / simulate the experimental d\textit{I}/d\textit{V} spectra reported in Figure 5. Spectra in Fig. 5(c) were fitted,  while for the spectra in (f), fitting was not possible and a fine-tuning approach was employed.
}
\end{table}

\newpage

\subsection{CAS calculations}
The starting point for the CAS calculations is a tight-binding model where we only consider the $p_z$ orbitals of the carbon atoms that compose the nanographenes.
We consider a typical value for the first neighbor hopping, $t_1 = -2.7$~eV.
Moreover, we take a third neighbor hopping $t_3 = 0.1 t_1$, in line with previous work and essential to capture the magnetic properties of triangulene lattices~\cite{ortiz_theory_2023}. 
The corresponding Hamiltonian reads as
\begin{equation}
\mathcal{\hat{H}}_0 = t_1 \sum_{\sigma} \sum_{\langle i,j \rangle } \hat{c}^\dagger_{i,\sigma} \hat{c}_{j,\sigma} + t_3 \sum_{\sigma} \sum_{\langle   i,j \rangle  } \hat{c}^\dagger_{i,\sigma} \hat{c}_{j,\sigma},
\end{equation}
where $\hat{c}_{i,\sigma}$ denotes the annihilation operator for an electron in carbon site $i$ with spin $\sigma = \uparrow, \downarrow$.
This single-particle Hamiltonian can be easily diagonalized numerically, leading to a set of molecular orbitals.
The Fermi level $\mu$ is defined assuming nanographenes at charge neutrality, i.e., one electron per carbon site.

In our CAS calculations, we first choose an active space of molecular orbitals close to the Fermi level. 
The underlying approximation is to assume that all molecular orbitals with energy below (above) the active space are doubly occupied (empty).
Then, we include interactions in the Hubbard form,
\begin{equation}
\mathcal{\hat{H}}_U = U \sum_i \hat{n}_{i,\uparrow} \hat{n}_{i,\downarrow},
\end{equation}
where $U>0$ is the on-site Hubbard repulsion, taken as a free parameter, and $\hat{n}_{i,\sigma} = \hat{c}^\dagger_{i,\sigma} \hat{c}_{i,\sigma}$.
Finally, the total (many-body) Hamiltonian $\mathcal{\hat{H}} = \mathcal{\hat{H}}_0 + \mathcal{\hat{H}}_U$ is represented in a restricted basis set---where we consider all multielectronic configurations that can be obtained with $N_e$ electrons in the previously selected $N_{MO}$ molecular orbitals---and diagonalized numerically.
This is referred to as the CAS($N_e$,$N_{MO}$) approximation.
Importantly, given that we consider nanographenes described by a bipartite lattice at half filling, $\mathcal{\hat{H}}_0$ features molecular orbitals with symmetric energy with respect to $\mu = 0$.
Therefore, our active space contains all (singly-occupied) zero-energy states, plus an equal number of molecular orbitals with energy above and below $\mu$, which implies $N_e = N_{MO}$.

\begin{figure*}[b!]
\begin{center}
\includegraphics[width=\columnwidth]{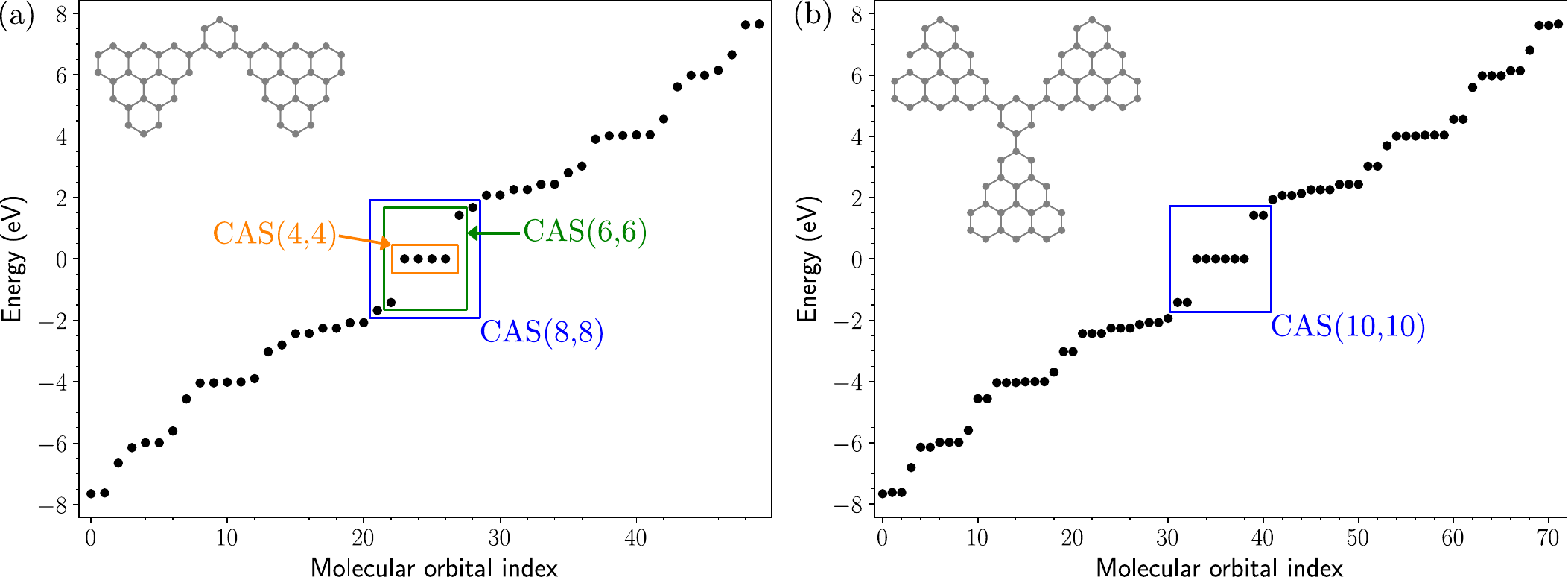}
\caption{
Energy levels, obtained with the tight-binding model, for (a) ferromagnetic triangulene dimer and (b) ferromagnetic triangulene trimer.
Solid black lines denote the Fermi level.
Boxes represent the different choices of active space used in the CAS calculations of Fig.~\ref{fig:CAS2}.
}
\label{fig:CAS1}
\end{center}
\end{figure*}

In Fig.~\ref{fig:CAS1}, we show the (single-particle) tight-binding energy levels of ferromagnetic triangulene dimers and trimers.
Given that benzene spacers link triangulenes via the same sublattice, dimers (trimers) have a sublattice imbalance of 4 (6), which implies the existence of the observed 4 (6) zero-energy states~\cite{sutherland_localization_1986}.
These singly-occupied zero-energy states constitute the minimal active space that can be used in our calculations, namely CAS(4,4) for dimers and CAS(6,6) for trimers.
Going beyond the minimal active space---up to CAS(10,10) due to memory constraints---allows to include corrections due to higher-energy molecular orbitals, although convergence of such Coulomb-driven exchange contributions is known to be problematic~\cite{jacob_theory_2022}.
Therefore, our CAS calculations do not aim at providing a quantitative description of the experimental results (namely the value of the exchange coupling), but rather to certify the picture of a ferromagnetic spin model.

\begin{figure*}[t!]
\begin{center}
\includegraphics[width=\columnwidth]{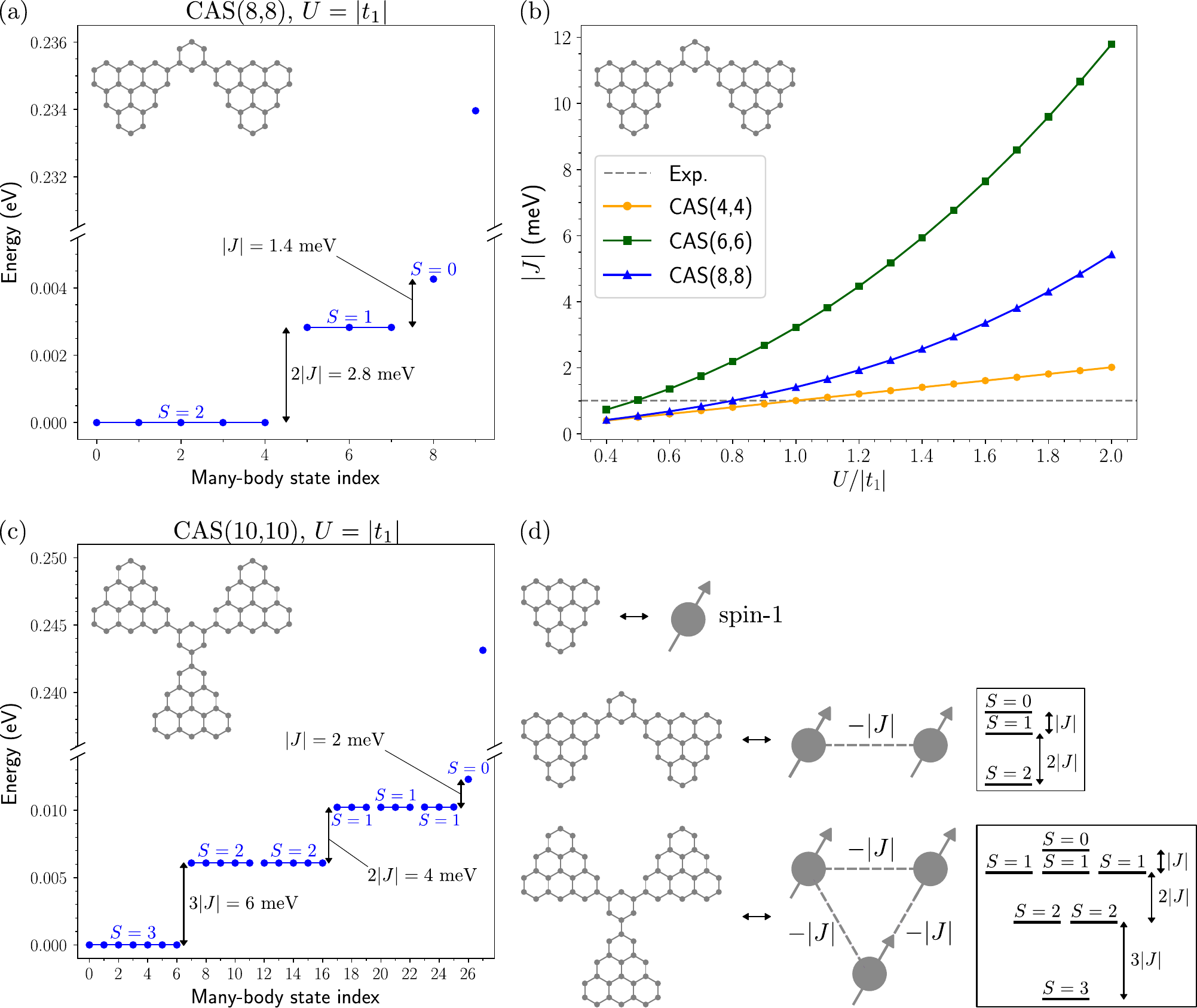}
\caption {
CAS calculations and mapping to the ferromagnetic spin model.
(a) Many-body energy levels, obtained with CAS(8,8) and $U=|t_1|$, for the ferromagnetic triangulene dimer. 
(b) Exchange coupling, obtained as half of the quintet-triplet splitting of the ferromagnetic triangulene dimer, as a function of $U$, for different choices of active space.
The gray dashed line denotes the experimental value.
(c) Many-body energy levels, obtained with CAS(10,10) and $U=|t_1|$, for the ferromagnetic triangulene trimer.
(d) Scheme illustrating the mapping between triangulenes coupled via same-sublattice benzene spacers and ferromagnetic spin-1 Heisenberg models.
The agreement between the energy levels of the spin models, shown in the right panels, and the low energy manifold of the CAS calculations (a,c), validates the use of a ferromagnetic spin-1 Heisenberg Hamiltonian and allows the theoretical determination of the exchange coupling constant $|J|$.
}
\label{fig:CAS2}
\end{center}
\end{figure*}

In Fig.~\ref{fig:CAS2}a, we show our CAS calculations for the ferromagnetic triangulene dimer, considering the largest affordable active space for this system\footnote{Note that degeneracies forbid the use of CAS(10,10).} and taking $U=|t_1|$ as a reference value, usual for nanographenes~\cite{turco_observation_2023,catarina_broken-symmetry_2023}.
We observe a low-energy manifold, composed of 9 states, well separated from the remaining higher-energy states.
Importantly, this low-energy manifold is compatible with the energy levels of a ferromagnetic spin-1 Heisenberg dimer (Fig.~\ref{fig:CAS2}d), sharing the same degeneracy pattern ($S=2$ ground state, followed by $S=1$ and $S=0$ excited states) and the same ratio of excitation energies (quintet-triplet excitation energy being twice as large as the triplet-singlet).
Analogous conclusions are drawn for the trimer, as shown in Fig.~\ref{fig:CAS2}c.
Therefore, we justify the use of a ferromagnetic spin-1 Heisenberg model to describe these systems.
Finally, regarding the magnitude of the exchange coupling constant $|J|$, Fig.~\ref{fig:CAS2}b shows that, despite significant oscillations with respect to the choice of the active space, our CAS calculations with $U \sim |t_1|$ are in reasonable agreement with experiments, where ferromagnetic exchange couplings of 1 meV were inferred from IETS.

\clearpage
\section{Supporting STM, STS and theoretical data}

\subsection{Precursors \textbf{2p} and \textbf{3p} as deposited on Au(111)}

\begin{figure*}[h]
\begin{center}
\includegraphics[width=17.4cm]{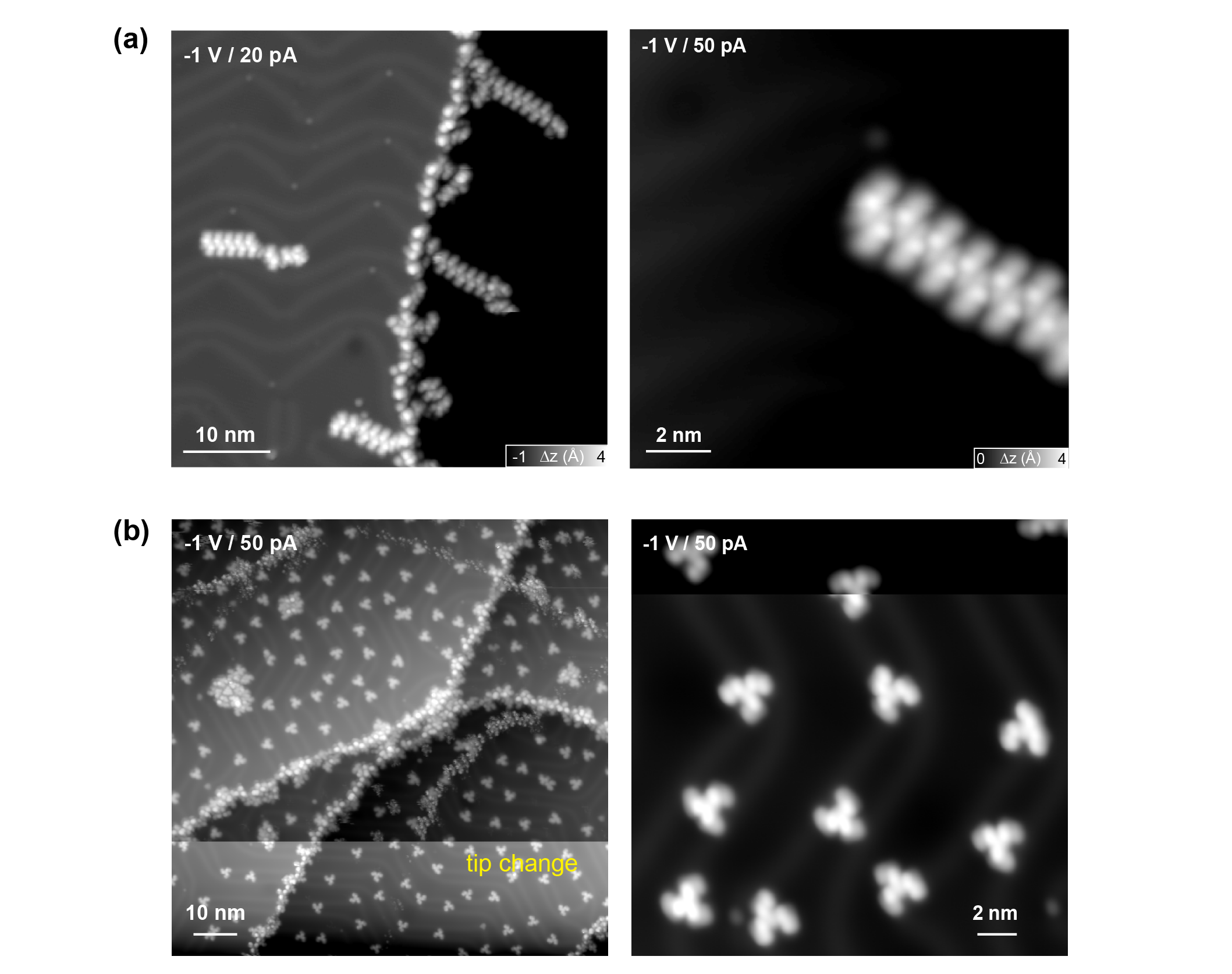}
\caption {Overview STM images of sub-monolayer coverages of \textbf{2p} (a) and \textbf{3p} (b) deposited on a Au(111) surface held at room temperature.}
\label{Fig_RTphase}
\end{center}
\end{figure*}

\newpage
\subsection{STS with carbon monoxide (CO) functionalized tip}
\begin{figure*}[h]
\begin{center}
\includegraphics[width=17.4cm]{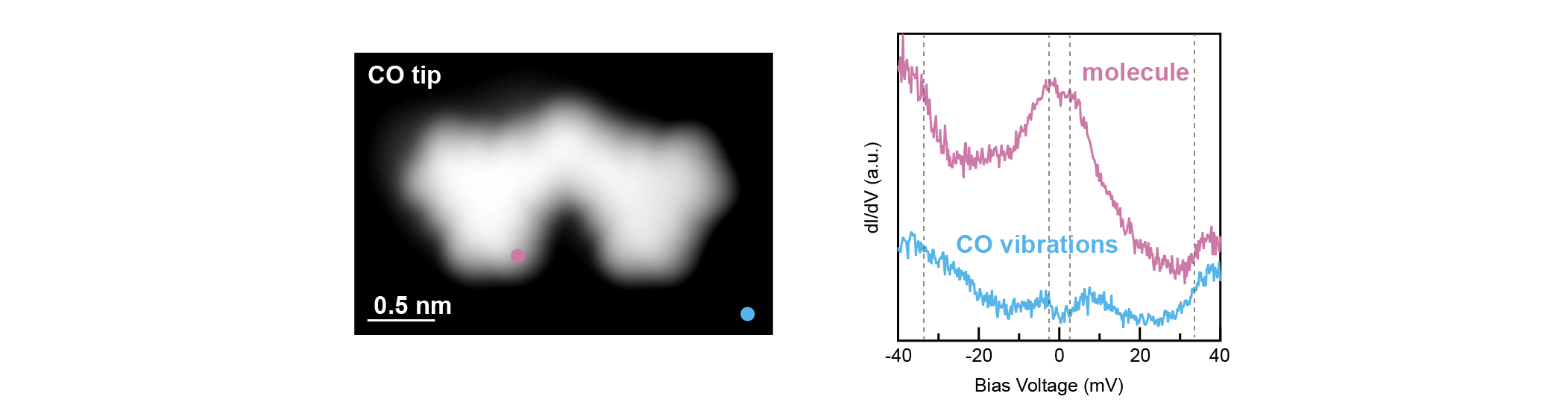}
\caption {HR-STS spectroscopy with CO-functionalized tip acquired on molecule \textbf{2} and on the bare Au(111) surface. The spectra reveal the presence of vibrational inelastic excitations, which can be misinterpreted as spin-excitation steps. Open feedback parameters \didv spectra: (c) $V=-50$ mV, $I=800$ pA. Lock-in modulation $V_{rms}=700 $ $\mu$V.}
\label{Fig_COvibrations}
\end{center}
\end{figure*}

\subsection{Trimer's electronic characterization}
\begin{figure*}[h]
\begin{center}
\includegraphics[width=17.4cm]{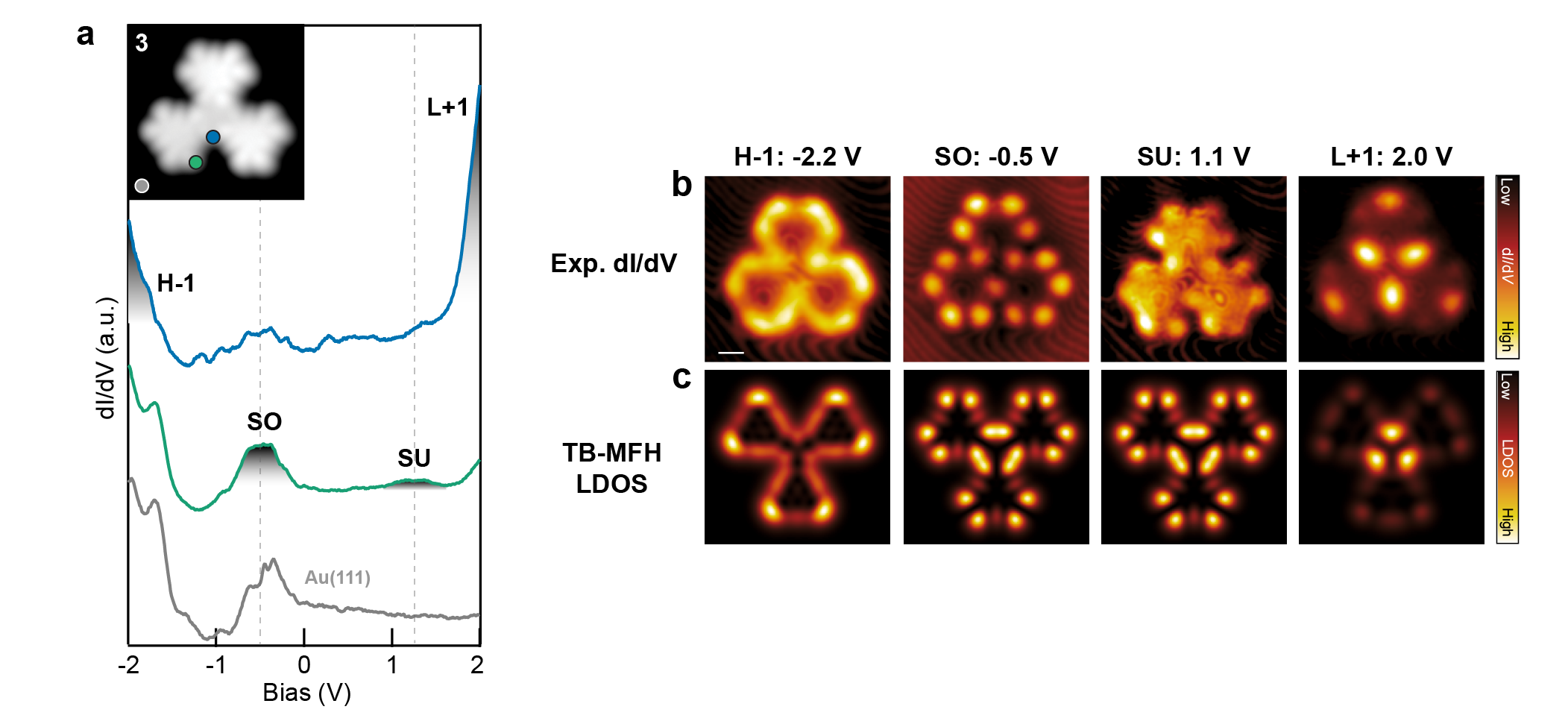}
\caption{Experimental electronic of \textbf{3}. (a) \didv spectroscopy acquired with a metal tip on three different locations (marked with filled circles) on molecule \textbf{3}, revealing four distinct molecular orbital (MO) resonances (open feedback parameters: $V=-2.0$ V, $I=350$ pA; $V_{rms}=18 $ mV). (b) Constant-current \didv spatial mapping of the detected MO resonances ( $I=300$ pA; $V_{rms}=19$ mV), along with the corresponding MFH-LDOS maps (c). Notably, the resonances labelled as H-1 and L+1 represent a superposition of higher-order occupied and empty molecular orbitals.}
\label{Electronic_charact_trimer}
\end{center}
\end{figure*}

\newpage

\section{Synthetic procedures}

\subsection{General Methods and Materials}
Unless otherwise noted, commercially available starting materials, reagents, catalysts, and dry solvents were used without further purification. Reactions were performed using standard vacuum-line and Schlenk techniques. All starting materials were obtained from TCI, Sigma Aldrich, abcr, Alfa Aesar, Acros Organics, or Fluorochem. The catalysts were purchased from Strem. Column chromatography was performed on silica (SiO\textsubscript{2}, particle size 0.063-0.200 mm, purchased from VWR). Silica-coated aluminum sheets with a fluorescence indicator (TLC silica gel 60 F254, purchased from Merck KGaA) were used for thin-layer chromatography. Dichloromethane-d\textsubscript{2} (99.9 atom \% D) and 1,1,2,2-Tetrachloroethane-d\textsubscript{2} ( $\geq$99.5 atom \% D) were purchased from Sigma Aldrich. The key building block 9-(4-bromo-2,6-dimethylphenyl)anthracene (\textbf{4}) was synthesized through five-step procedures in our previous work \cite{mishra_collective_2020}.

NMR data were recorded on a Bruker AV-II 300 spectrometer operating at 300 MHz for \textsuperscript{1}H and 75 MHz for \textsuperscript{13}C. Measurements were made at room temperature (296 K; AV-II 300) unless otherwise stated. Chemical shifts ($\delta$) are reported in ppm. The coupling constants (\textit{J}) are reported in Hz. Dichloromethane-d\textsubscript{2} ($\delta$(\textsuperscript{1}H) = 5.32 ppm,\textit{ $\delta$}(\textsuperscript{13}C) = 53.8 ppm) and 1,1,2,2-Tetrachloroethane-d\textsubscript{2} were used as solvent. The following abbreviations are used to describe peak patterns as appropriate: \textit{s} = singlet, \textit{d} = doublet, \textit{t} = triplet, \textit{q} = quartet, and \textit{m} = multiplet. High-resolution matrix-assisted laser desorption/ionization time-of-flight (MALDI-TOF) MS was recorded on a Bruker Autoflex Speed MALDI-TOF MS (Bruker Daltonics, Bremen, Germany). All of the samples, were prepared by mixing the analyte and the matrix, 1,8-dihydroxyanthracen-9(10\textit{H})-one (dithranol, purchased from Fluka Analytical, purity $>$ 98\%) or \textit{trans}-2-[3-(4-tert-butylphenyl)-2-methyl-2-propenylidene]malononitrile (DCTB, purchased from Sigma Aldrich, purity $>$ 99\%) in the solid state.

\subsection{Detailed Synthetic Procedures}
\begin{figure*}[h]
\begin{center}
\includegraphics[width=17.4cm]{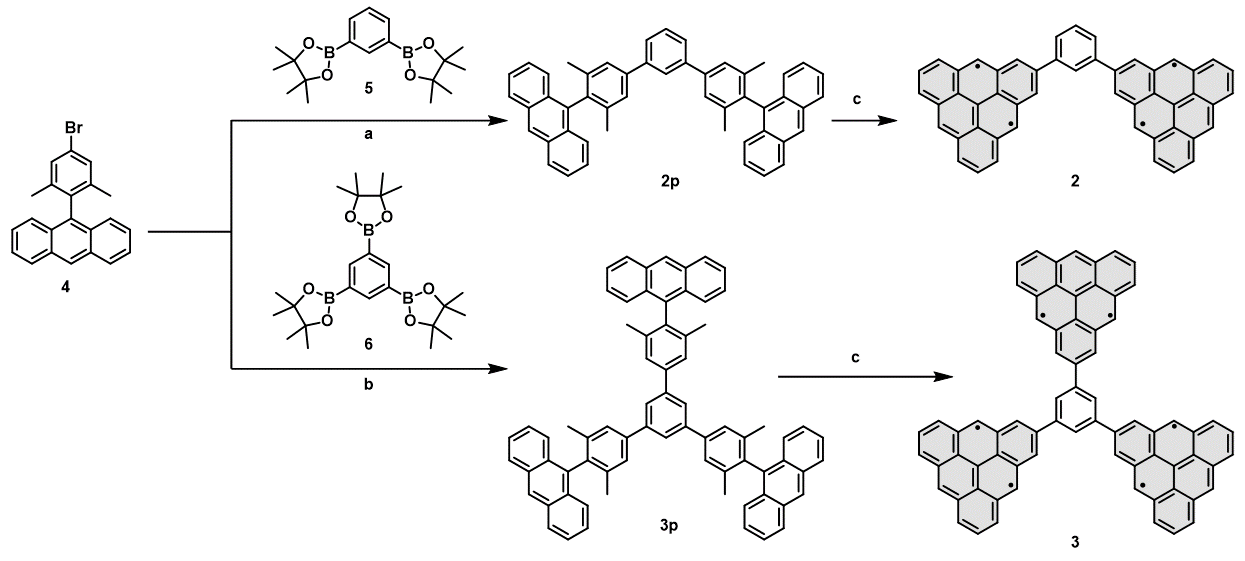}
\caption{Synthetic route to ferromagnetic triangulene dimer \textbf{2} and ferromagnetic triangulene trimer \textbf{3}. Reagents and conditions: (a) PdCl$_{2}$(dppf)CH$_{2}$Cl$_{2}$, K$_{3}$PO$_{4}$, dioxane, 100 °C, 16 h, 34\%. (b) PdCl$_{2}$(dppf)CH$_{2}$Cl$_{2}$, K$_{3}$PO$_{4}$, dioxane, 85 °C, 16 h, 65\%. (c) Au (111) held at 310 °C and 300 °C for \textbf{2} and \textbf{3}, respectively. }
\label{Figure S7 Synthetic route}
\end{center}
\end{figure*}
As shown in Figure S7, a Suzuki-coupling reaction of \textbf{4} with commercially available 1,3-bis(4,4,5,5-tetramethyl-1,3,2-dioxaborolan-2-yl)benzene (\textbf{5}) or 1,3,5-tris(4,4,5,5-tetramethyl-1,3,2-dioxaborolan-2-yl)benzene (\textbf{6}) gave precursors \textbf{2p }and\textbf{ 3p} in 34\% yield and 65\% yield, respectively. Subsequently, these two precursors were applied for the on-surface synthesis of triangulene dimer (\textbf{2}) and trimer (\textbf{3}). 

\newpage
\textbf{Synthesis of 9,9'-(3,3'',5,5''-tetramethyl-[1,1':3',1''-terphenyl]-4,4''-diyl)dianthracene (2p)}:
\begin{figure*}[h]
\begin{center}
\includegraphics[width=17.4cm]{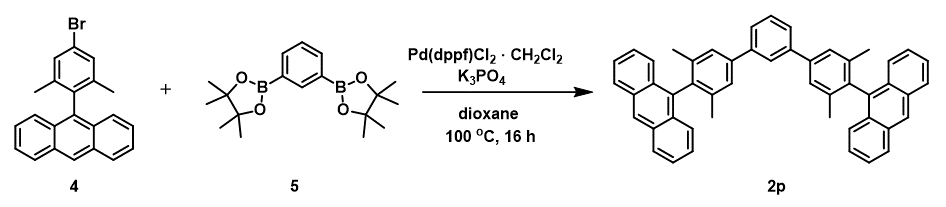}
\label{Synthesis of 2p}
\end{center}
\end{figure*}

A 25 mL Schlenk tube charged with 9-(4-bromo-2,6-dimethylphenyl)anthracene (\textbf{4})\textsuperscript{[6]} (50 mg, 0.138 mmol), the commercially available 1,3-bis(4,4,5,5-tetramethyl-1,3,2-dioxaborolan-2-yl)benzene (\textbf{5}) (18.3 mg, 0.0554 mmol), [1,1'-bis(diphenylphosphino)ferrocene]dichloropalladium(II) complex with dichloromethane (Pd(dppf)Cl\textsubscript{2}\textbf{·}CH\textsubscript{2}Cl\textsubscript{2}) (11.3 mg, 0.0138 mmol) and K\textsubscript{3}PO\textsubscript{4} (176 mg, 0.830 mmol) was evacuated and charged with argon three times. Then degassed 1,4-dioxane (10 mL) was added and the reaction mixture was stirred at 100 °C for 36 h under an argon atmosphere. After cooling to room temperature, the mixture was poured into water and extracted with dichloromethane (DCM) three times. The organic layer was washed with brine and dried over anhydrous MgSO\textsubscript{4}. The solvent was removed under vacuum. The residue was purified by silica gel column chromatography (iso-hexane: DCM = 5:1) to give compound \textbf{2p} as a white solid (30 mg, 34\%). \textsuperscript{1}H NMR (300 MHz, CD\textsubscript{2}Cl\textsubscript{2}) $\delta$ (ppm) = 8.56 (s, 2H), 8.20 - 8.07 (m, 5H), 7.85 - 7.76 (m, 2H), 7.68 (s, 5H), 7.59 - 7.47 (m, 8H), 7.39 (ddd, \textit{J} = 8.7, 6.5, 1.3 Hz, 4H), 1.85 (s, 12H). \textsuperscript{13}C NMR (75 MHz, CD\textsubscript{2}Cl\textsubscript{2}) $\delta$ 142.22, 140.99, 138.85, 137.42, 135.82, 132.26, 130.18, 129.91, 129.24, 126.89, 126.87, 126.52, 126.46, 126.40, 126.31, 125.86, 20.50. HR-MALDI-TOF (matrix: DCTB): calc. for [M]\textsuperscript{+}: 638.2969, found for [M]\textsuperscript{+}: 638.2963 (deviation: -0.94 ppm).

\textbf{Synthesis of 9,9'-(5'-(4-(anthracen-9-yl)-3,5-dimethylphenyl)-3,3'',5,5''-tetramethyl
-[1,1':3',1''-terphenyl]-4,4''-diyl)dianthracene (3p)}:

\begin{figure*}[h]
\begin{center}
\includegraphics[width=17.4cm]{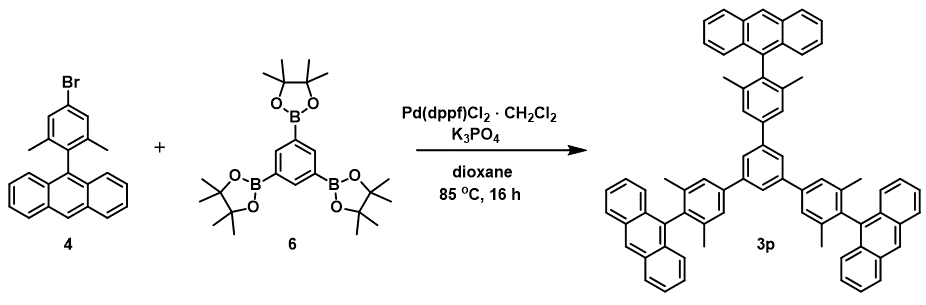}
\label{Synthesis of 3p}
\end{center}
\end{figure*}

A 25 mL Schlenk tube charged with compound \textbf{4}\textsuperscript{[6]} (26.2 mg, 0.072 mmol), the commercially available \textbf{6} (10 mg, 0.022 mmol), [1,1'-bis(diphenylphosphino)ferrocene]dichloropalladium(II) complex with dichloromethane (Pd(dppf)Cl\textsubscript{2}\textbf{·}CH\textsubscript{2}Cl\textsubscript{2}) (3.6 mg, 20\% mmol) and K\textsubscript{3}PO\textsubscript{4} (42 mg, 0.20 mmol) was evacuated and charged with argon three times. Then degassed 1,4-dioxane (8 mL) was added and the reaction mixture was stirred at 85 ° C for 16 h under an argon atmosphere. After cooling to room temperature, the mixture was poured into water and extracted with dichloromethane (DCM) three times. The organic layer was washed with brine and dried over anhydrous MgSO\textsubscript{4}. The solvent was removed under vacuum. The residue was purified by silica gel column chromatography (iso-hexane: DCM = 3:1) to give \textbf{3p} as a white solid (13 mg, 65\%). \textsuperscript{1}H NMR (300 MHz, C\textsubscript{2}D\textsubscript{2}Cl\textsubscript{4}-\textit{d}\textsubscript{2}) $\delta$ 8.49 (s, 3H), 8.12 (s, 3H), 8.06 (d, \textit{J} = 8.4 Hz, 6H), 7.71 (s, 6H), 7.57 – 7.52 (m, 6H), 7.46 (ddd, \textit{J} = 8.3, 6.6, 1.3 Hz, 6H), 7.35 (ddd, \textit{J} = 8.7, 6.5, 1.3 Hz, 6H), 1.85 (s, 18H). \textsuperscript{13}C NMR (76 MHz, C\textsubscript{2}D\textsubscript{2}Cl\textsubscript{4}-\textit{d}\textsubscript{2}) $\delta$ 142.27, 140.38, 138.34, 136.91, 135.35, 131.57, 129.60, 128.73, 126.55, 125.98, 125.89, 125.40, 125.07, 120.35, 20.47. HR-MALDI-TOF (matrix: DCTB): calc. for [M]\textsuperscript{+}: 918.4220, found for [M]\textsuperscript{+}: 918.4221 (deviation: -0.1 ppm).

\newpage
\begin{figure*}[h]
\begin{center}
\includegraphics[width=17.4cm]{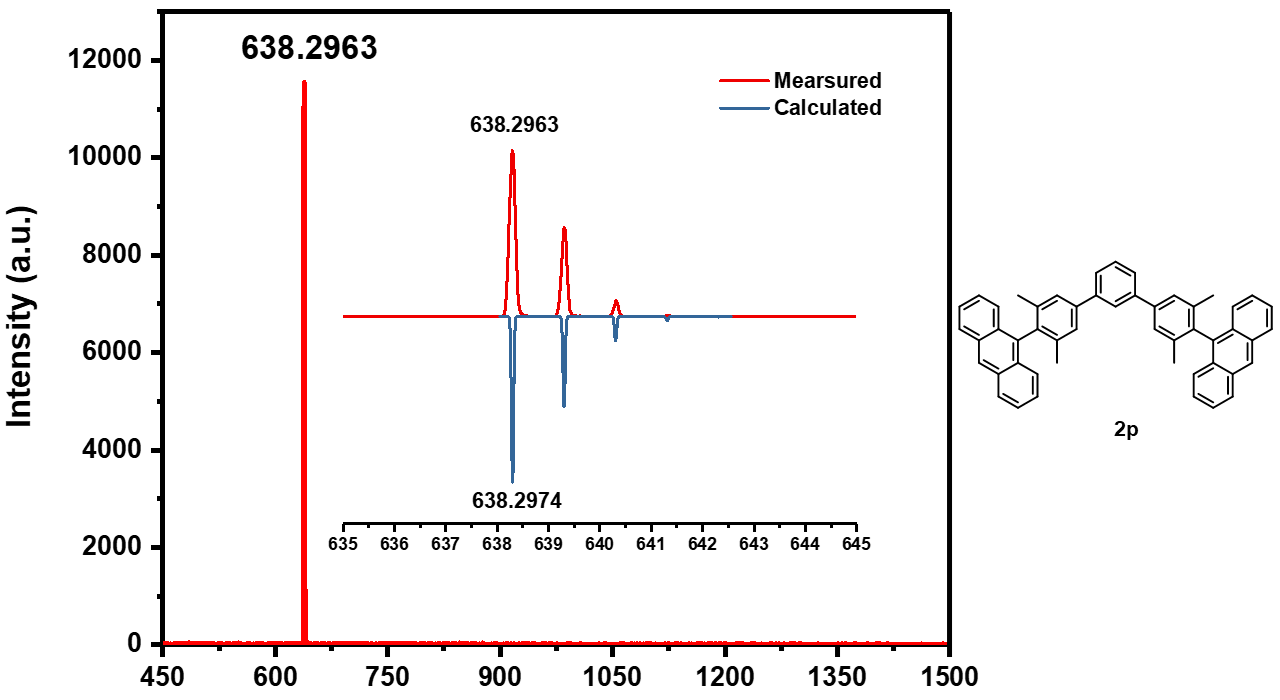}
\caption {Liquid-state HR-MALDI-TOF-MS of \textbf{2p} (matrix: DCTB)}
\label{Figure S8 HR MALDI-TOF MS spectrum of 2p}
\end{center}
\end{figure*}
\begin{figure*}[h]
\begin{center}
\includegraphics[width=17.4cm]{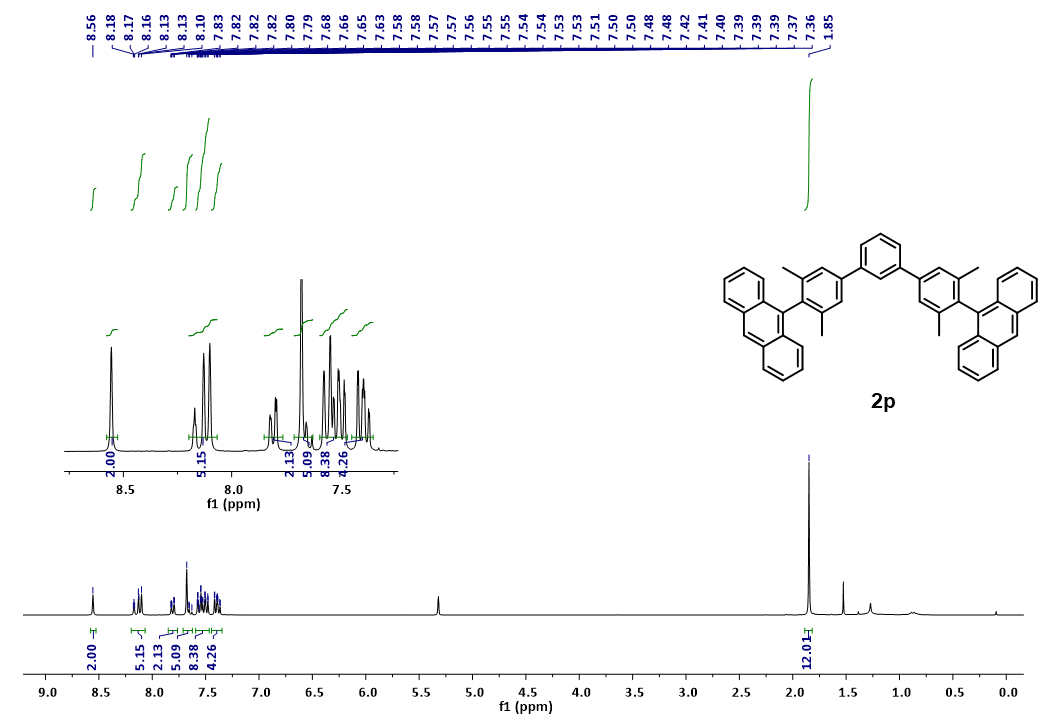}
\caption {\textsuperscript{1}H-NMR spectrum of \textbf{2p} dissolved in CD\textsubscript{2}Cl\textsubscript{2}, 300 MHz, 296 K. }
\label{Figure S9 HNMR spectrum of 2p}
\end{center}
\end{figure*}

\newpage
\begin{figure*}[h]
\begin{center}
\includegraphics[width=17.4cm]{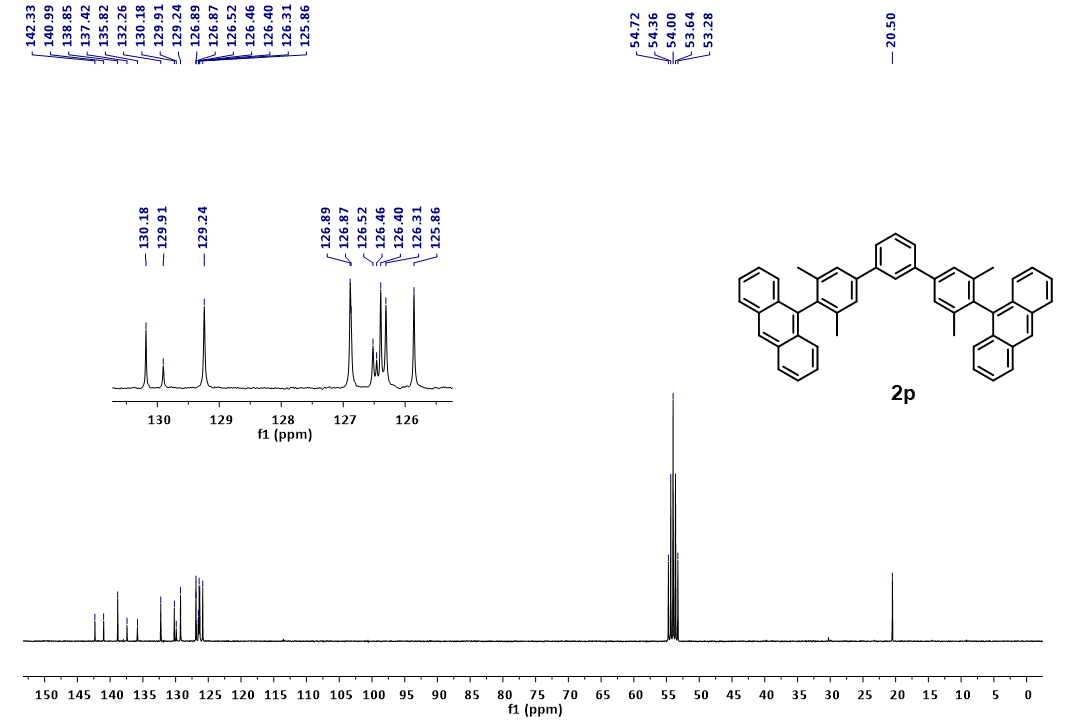}
\caption {\textsuperscript{13}C-NMR spectrum of \textbf{2p} dissolved in CD\textsubscript{2}Cl\textsubscript{2}, 75 MHz, 296 K. }
\label{Figure S10 CNMR spectrum of 2p}
\end{center}
\end{figure*}

\begin{figure*}[h]
\begin{center}
\includegraphics[width=17.4cm]{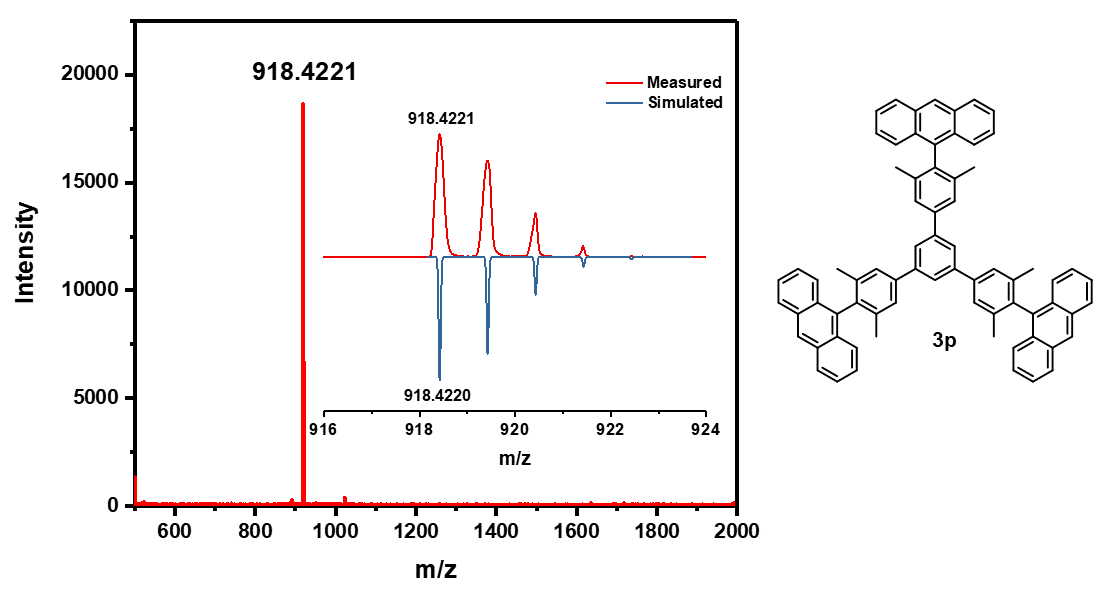}
\caption {Liquid-state HR-MALDI-TOF-MS of \textbf{3p} (matrix: DCTB)}
\label{Figure S11 HR MALDI-TOF MS spectrum of 3p}
\end{center}
\end{figure*}

\newpage
\begin{figure*}[h]
\begin{center}
\includegraphics[width=15.8cm]{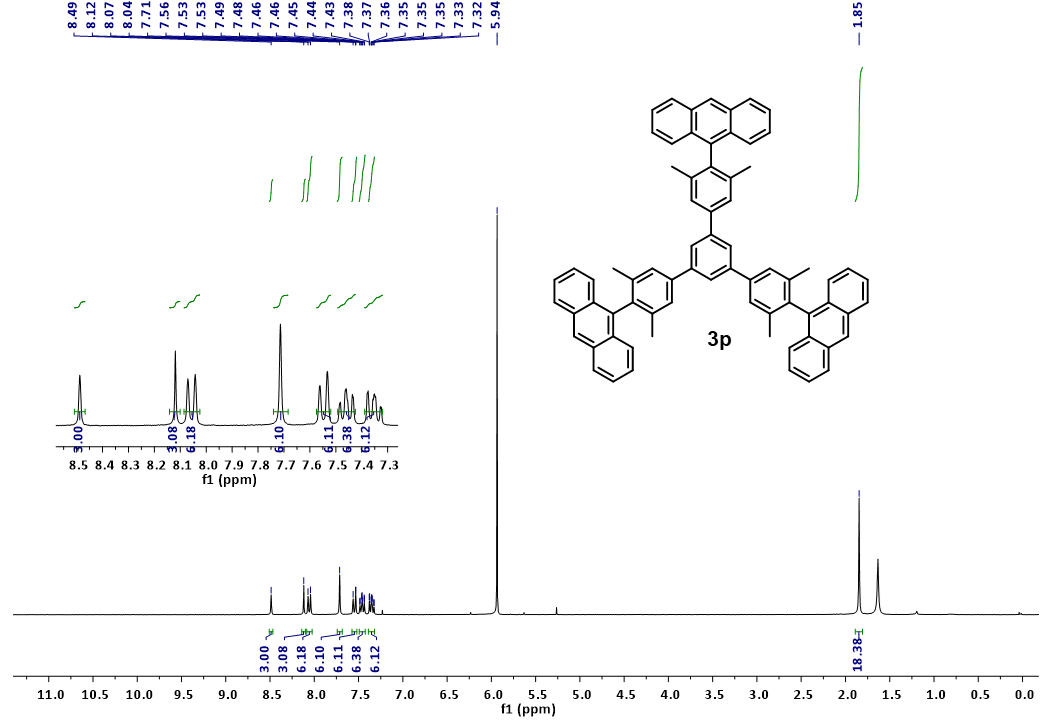}
\caption {\textsuperscript{1}H-NMR spectrum of \textbf{3p} dissolved in C\textsubscript{2}D\textsubscript{2}Cl\textsubscript{4}-\textit{d}\textsubscript{2}, 300 MHz, 296 K}
\label{Figure S12 HNMR spectrum of 3p}
\end{center}
\end{figure*}
\begin{figure*}[h]
\begin{center}
\includegraphics[width=15.8cm]{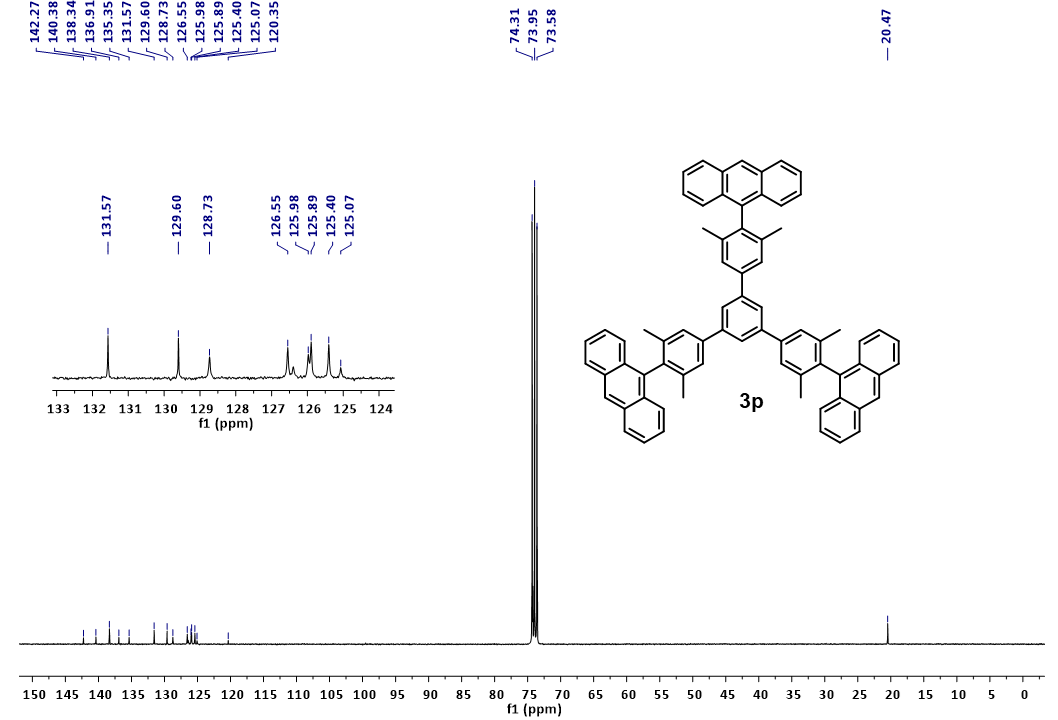}
\caption {\textsuperscript{13}C-NMR spectrum of \textbf{3p} dissolved in C\textsubscript{2}D\textsubscript{2}Cl\textsubscript{4}-\textit{d}\textsubscript{2}, 75 MHz, 296 K.}
\label{Figure S13 CNMR spectrum of 3p}
\end{center}
\end{figure*}

%%%%%%%%%%%%%%%%%%%%%%%%%%%%%%%%%%%%%%%%%%%%%%%%%%%%%%%%%%

\end{document}